\documentclass[authoryear]{amsart}
\usepackage{amsmath,lscape}
\usepackage{natbib}
\usepackage{amssymb}
\usepackage{wasysym}
\usepackage{dcolumn}
\usepackage{pdflscape}
\usepackage[T1]{fontenc}
\usepackage{attrib}
\usepackage{a4wide}
\usepackage{supertabular}
\usepackage{multirow}
\usepackage{setspace} % alterar espaamento entre linhas
\usepackage{graphicx,color} % Para escrever em cores
\usepackage{graphics}
\usepackage{subfig}
\usepackage{longtable}
\usepackage{ae}
\usepackage{rotating}
\usepackage{url}
\usepackage{morefloats}
\usepackage{gensymb}
\usepackage{bm}

\usepackage{hyperref} % para as citaes e refs voltarem a cor normal comentar esta parte
\hypersetup{
  colorlinks = true,
  linkcolor=blue,   % color of internal links
  citecolor=blue,   % color of links to bibliography
  urlcolor=blue,    % color of external links
  pagebackref=true,
  implicit=false,
  bookmarks=true,
  bookmarksopen=true,
  pdfdisplaydoctitle=true
}
\begin{document}

\title{A continuous spatio-temporal approach to estimate climate change}

%\cortext[cont]{Corresponding author -  Av. dos Bandeirantes 3900,  14040-905, Ribeir\~ao Preto, SP, Brazil.   Tel.: +55-16-33290867 -  email - mplaurini@gmail.com}
\author{M\'arcio Poletti  Laurini}%\corref{cont}
\address{Department of Economics - FEA-RP University of São Paulo. -  Av. dos Bandeirantes 3900,  14040-905, Ribeir\~ao Preto, SP, Brazil.   Tel.: +55-16-33290867 -  email - mplaurini@gmail.com}

\begin{abstract}

We introduce a method for decomposition of trend, cycle and seasonal components in spatio-temporal models and apply it to investigate the existence of climate changes in temperature and rainfall series. The method incorporates critical features in the analysis of climatic problems - the importance of spatial heterogeneity, information from a large number of weather stations, and the presence of missing data. The spatial component is based on continuous projections of spatial covariance functions, allowing modeling the complex patterns of dependence observed in climatic data. 
 We apply this method to study climate changes in the Northeast region of Brazil, characterized by a great wealth of climates and large amplitudes of temperatures and rainfall. The results show the presence of a tendency for temperature increases, indicating changes in the climatic patterns in this region. 
\end{abstract}

\maketitle

%\doublespacing

\section{Introduction\label{intro}}

There is an important discussion in society (e.g., \cite{Howe2013})  about the patterns of climate change observed in recent years, whether these are caused by human actions (\cite{Kaufmann19072011}) or are the result of natural trends, and the impact of these changes on the environment and economy. A key issue in this problem is the correct measurement of climate change patterns, and especially if there are changes in trends and cyclical patterns of climatic measures such as temperature and rainfall. 

Measuring changes in climate patterns is a complex problem, since short- and long-term climate patterns are determined by a combination of factors, such as the absorption of solar radiation, sea temperature, air mass flow and clouds, as well as geographic aspects such as altitude and proximity to the ocean. Another fundamental difficulty is that the dynamics of these factors and their relations are phenomena of high complexity, requiring the estimation and simulation of theoretical models with large computational cost.

One of the ways to verify the existence of changes in climate patterns is through the estimation of trends and periodic components in climate-related time series models, as in \cite{Bloomfield1992},  \cite{Gordon1991}, \cite{Zheng1999}, \cite{ Kaufmann2006} and  \cite{Proietti2017}. In these works, the objective is to decompose the temporal variability observed in climatic series using statistical methods for the extraction of trends, seasonal and cycle components. In this approach, climate changes are identified as changes in long-term trends or in the observed seasonal pattern, as in \cite{Proietti2017}. These decompositions allow a simple statistical interpretation of climate change patterns, summarizing the wealth of information contained in these data. 

Note that this formulation is useful to detect climate changes, since it allows the estimation of variations in the behavior of the series. Traditional methods for climate modeling, such as the use of regressions of temperature as function of fixed measurements, such as altitude and geographic coordinates (e.g. \cite{Stape2013b}), are invariant over time and thus do not allow estimation of changes in trend and periodic components.

Inference procedures on climate patterns also face some unusual problems in terms of statistical and econometric methods. The first difficulty is associated with the available sources of information, usually based on a large number of monitoring stations. Although the existence of many sources of climate information is an advantageous feature by allowing greater accuracy in the inference processes, the statistical methods used to extract trends and cycles are usually not adapted to this feature.

Most of the available methods of time series decomposition into trend, seasonality and cycle components, also known as structural time series decompositions (e.g., \cite{Harvey1989}), are not fully adapted to the data sources used in climatology. These data sources present three particularly difficult problems for the analysis of time series. The first is the dimensionality of the data. As already commented above, the data come from a wide network of weather stations, generating a large number of time series, e.g., \cite{Storelvmo2016}. The econometric methods of trend, seasonal and cycle extractions are usually based on univariate or low-dimensional models, due to computational difficulties such as the numerical maximization of likelihood functions with a large number of parameters and latent factors. In this way, most trend and cycle extraction studies are based on univariate analyses or else they apply some dimension reduction procedures, such as principal component analysis or data aggregation. 

The second difficulty is the importance of the spatial distribution of climatic effects, reflected in the statistical analysis of the values observed by the network of monitoring stations. In this respect, the measurement of climatic patterns is in fact a large space-time problem. Again, the usual econometric methods used in structural time series decompositions, such as Kalman filter-based methods, fail to properly incorporate spatial effects. The use of these econometric methods in this problem again usually implies some form of reduction of the existing complexity, such as data aggregation, for example, by analyzing temperature averages for a region.

There is also a practical difficulty related to the patterns of operation of weather monitoring stations. Many weather stations are operated automatically, and they are often located in regions with difficult access. In the event of an operational problem, these stations can become inoperative, failing to collect data. There is also the addition and replacement of stations in new regions. In statistical terms, this means the time series collected by these stations will have a significant portion of missing data, which adds a new inference problem, leading to the need for interpolation or data imputation methods, which can affect the measurements of the trend, seasonality and cycle components related to climate measures of interest. 

In this work, we will introduce a method that allows decomposition of climatic series into trend, seasonal and cycle components, but also incorporates the spatial heterogeneity and high dimensionality that exists in these data, and allows dealing with missing data characteristics in these series without the need for initial interpolation or data imputation procedures. The method combines elements of structural time series decompositions (e.g., \cite{Harvey1989} and  \cite{Proietti1991}) with the spatio-temporal models with continuous spatial random effects, based on the equivalence between the solution of a stochastic partial differential equation and spatial covariance functions, introduced by \cite{spde}. In the method proposed in this article, the time series are decomposed into common components of trend, seasonality and cycle, similar to a time series structural decomposition, but the innovation process in each location contains an error component projected in the spatial continuum, characterized by a spatial covariance function of the Matérn class. This formulation can be thought of as a process of decomposing geostatistical time series (e.g.,  \cite{Cressie1993} and  \cite{Cressie2011} into a sum of trend, seasonal and cycle components plus the effect of additional covariates.

The decomposition used in this work can be represented by the following structure:

\begin{equation}
\begin{array}{c}
y\left(s,t\right)=\mu_t+s_t+c_t+z\left(s,t\right)\beta+\xi\left(s,t\right)+\epsilon\left(s,t\right)\\
\mu_t=\mu_{t-1}+\eta_\mu\\
s_t=s_{t-1}+s_{t-2}+...s_{t-m-1}+\eta_s\\
c_t=\phi_1 c_{t-1}+ \phi_2 c_{t-2}+\eta_c\\
\xi\left(s,t\right)=\omega\left(s,t\right)\\
Cov\left(\omega\left(s,t\right)\right)=\mathcal{C}\left(h\right)
\end{array}\label{model1}
\end{equation}

\noindent where $y\left(s,t\right)$  represents the observation $y$ at location $s$ and in period $t$, $\mu_t, s_t$ and $c_t$  are the components of trend, seasonality and cycle, with independent Gaussian innovation components $\eta_\mu$, $\eta_s$  and  $\eta_c$;    $z\left(s,t\right)$  is a set of covariates observed in the location $s$ and period $t$, $\epsilon\left(s,t\right)$, is an non spatial independent Gaussian innovation  component and $\xi\left(s,t\right)$  is a spatial random effect, represented by a process  $\omega\left(s,t\right)$  continuously projected in space and characterized by the spatial covariance function $\mathcal{C}\left(h\right)$, with  $h=||s-s^{\prime}||$  being the Euclidian distance between locations $s$ and  $s^{\prime}$, in this work modeled  as a covariance function of the Matérn class. This continuous projection is the main feature of this formulation, and its use allows solving the problems discussed above.

Since a continuous formulation is used to represent space, it is not necessary to have regular observations of each monitoring station. If a certain location is not observed at a certain time $t$, its value can be estimated through the continuous spatial covariance function, similar to a kriging process, now in a dynamic context. In this structure, we retrieve the specific values at each point in space by summing the estimated components of trend, cycle and seasonality with the continuous projection of the spatial random effect. In this form, both missing data and the other points in the continuum are treated as unobserved parameters, and estimated by the proposed model. Note that this structure deals simultaneously with the three problems discussed above. This representation incorporates a rich spatial structure in the model, allows recovering all points in the spatial continuum and solves the problems of high dimensionality and missing data through a decomposition of trend, seasonality, cycle and spatial random effect.

To perform the estimation of the general model given by Eq. (\ref{model1}), we use the powerful representation and inference structure introduced in \cite{spde}. In this work, the authors show that it is possible to represent a spatial covariance process in the continuum using a representation based on the equivalence between the solution of a certain class of stochastic differential partial equations (spde) and spatial covariance functions. This formulation allows representing continuous spatial processes using finite element methods, through base expansions, in a computationally efficient structure. The authors also show that this formulation can be associated with Gaussian Markov random fields, allowing the use of hierarchical structures and especially the use of Bayesian estimation methods based on integrated nested Laplace approximations, which are accurate analytical approximations for the posterior distribution of parameters and latent factors, introduced in \cite{Rue2009}. Our contribution is to combine this spatial representation with a Bayesian version of the basic structural model of \cite{Harvey1989}, using the fact that the additive structure of these processes also represents a Gaussian Markov random field, formulated as a hierarchical model, as discussed in \cite{Rue2010}. 

We use this structure to analyze the existence of changes in the climate patterns of the Northeast region of Brazil. This region is characterized by a huge diversity of climatic patterns, with great intra- and inter-annual variability in temperature and rainfall series. This richness of patterns in space and time makes this dataset a very interesting application for methods of extracting spatio-temporal climate changes. We analyze quarterly series of average and maximum temperatures and rainfall between 1961 and 2014 for a set of 91 weather monitoring stations in this region. As discussed above, this dataset presents the three fundamental problems associated with the econometric analysis of climatic series - high dimension (large number of time series provided by the weather stations), great spatial variability and missing data problems. The results indicate there is a trend of elevation in the average and maximum temperature series, and for the rainfall it is possible to observe a very relevant cyclical component. 

This work has the following structure - in the following Section \ref{method} we present the fundamental elements of the representation and inference structures used in this work. Section \ref{dataset} presents the analyzed dataset; Section \ref{results} shows the results obtained for the series of rainfall, average and maximum temperatures, Section \ref{extensions} presents some extensions of the method and finally Section \ref{conclusions} presents a discussion of the general results and conclusions.

\section{Methodology\label{method}}

The main innovation in this work is to combine a structure of trend, seasonality and cycles decomposition with the continuous spatial formulation presented in \cite{spde}. As this method is still of limited use in econometrics, we first give a basic description, following the presentations of \cite{spde}, \cite{Krainski2014} and \cite{Laurini2017}.

The method presented by   \cite{spde} is based on a Gaussian Markov random field (GMRF) representation of dynamic and spatial models (\cite{RueHeld2005}). The GMRF processes are represented by a structure of undirected graphs $\mathcal{G=}\{\mathcal{V},\mathcal{E}\}$, with $\mathcal{V}$ denoting a set of nodes and  $\mathcal{E}$    $\{i,j\}$ vertices, with $i,j$$\in \mathcal{V}$. A random vector process  $\boldsymbol{x}=(x_{1,}x_{2}, \ldots,x_{k})^{\top}\in\mathbb{R}^{n}$ represents a Gaussian Markov random field w.r.t. to the graph $\mathcal{G=}\{\mathcal{V},\mathcal{E}\}$,  with a  mean vector $E(\boldsymbol{x})=\boldsymbol{\mu}$   and precision matrix  $\boldsymbol{Q}$, iff its density can be written as  $\pi(\boldsymbol{x})=(2\pi)^{-n/2}|\boldsymbol{Q}|^{1/2}exp(-\frac{1}{2}(\boldsymbol{x}-\boldsymbol{\mu})^{\top}\boldsymbol{Q}(\boldsymbol{x}-\boldsymbol{u})$, with $Q_{ij}\neq0\Longleftrightarrow\{i,j\}\in\mathcal{G\vee}i\neq j$.  The GMRF structure can approximate a large set of processes, as well as the hierarchical formulation of several regression, generalized linear models, spatial and dynamic time series models, as discussed in  \cite{Rue2009}.

The continuous representation of spatial effects proposed in  \cite{spde}, is a computational representation of the results of  \cite{Whittle1954} and   \cite{Rozanov10973} on  the relationship between a class of spatial covariance process defined in a random field $x(u)$   and the solution of the  stochastic partial differential equation (spde) given by:

\begin{equation}
\left(\kappa-\Delta\right)^{\alpha/2}x(u)=W(u),\,\,\,\,\,,u\in\mathbb{R}^{d},\,\,\,\alpha=\nu+d/2,\,\,\kappa>0,\,\,\nu>0\label{eq:espde-1-1-1-1}
\end{equation}

\noindent  with  $\left(\kappa-\Delta\right)^{\alpha/2}$  being a  pseudo-differential operator and  $\Delta$ a Laplacian operator: 

\[
\Delta=\sum_{i=1}^{d}\frac{\partial^{2}}{\partial x_{i}^{2}}
\]

In this equation $\kappa$ is a spatial scale parameter and $\alpha$  and $\nu$  are parameters defining the smoothness of the realizations. The process $W(u)$  is a Gaussian white noise process, representing the spatial innovations. The main idea in  \cite{spde} is to use the equivalence relation to construct a computational representation of the continuous spatial covariance structure, using a finite elements representation of the equivalent spde  (e.g.  \cite{BrennerScott2007} and  \cite{Lord2014}) in some triangulated mesh, by using a basis expansion:

\begin{equation}
x(u)=\sum_{k=1}^{n}\psi_{k}(u)\omega_{k}\label{eq:basis-1-1-1-1-1}
\end{equation}

The solution is represented by a basis expansion $\psi_{k}(u)$ with weights  $\omega_{k}$  distributed as a Gaussian process,   with $n$  being the chosen number of vertices in the triangulated mesh. Usually the expansion is based on piecewise continuous functions inside each triangle. These weights are associated with the value of the random field process at the vertex of each triangle,  and inside the triangle the values are obtained by interpolation. The weight distribution is equivalent to the solution of the stochastic partial differential equation (\ref{eq:espde-1-1-1-1}) for a set of test functions $\phi_{k}$.

The two usual choices for the test functions are $\phi_{k}=\left(\kappa-\Delta\right)^{1/2}\psi_{k}$ for $\alpha=1$ and $\phi_{k}=\psi_{k}$ for $\alpha=2$, representing, respectively, least squares  and Galerkin solutions (e.g.  \cite{BrennerScott2007}). These functions determine the properties of the approximation to the true random field, and the use  of $\alpha=2$   imposes a  smoothness structure in the approximation of the random field, enabling better numerical properties in the computational representation of the continuous spatial process.  We also use this structure in our analysis.

This structure permits representing the spatial random effects, but it is possible to use very general structures for the conditional mean, using a hierarchical representation. Similar to the representation discussed in  \cite{Cameletti2013},  we represent a  continuously indexed  random field $Y(s,t)$ by:

\[
Y\left(s,t\right)=\{y\left(s,t)\right):\left(s,t\right)\in\mathcal{D}\subseteq\mathbb{R}^{2}\times\mathbb{R}
\]

\noindent  where  $s$   denotes a spatial coordinate and $t$  a time index. This model is characterized by spatial covariance function $Cov\left(\left(s,t\right)\left(s^{\prime},t^{\prime}\right)\right)=\sigma_\omega^{2}\mathcal{C}\left(h\right)$, for  $h=||s-s^{\prime}||$.  It is usually assumed that this covariance is spatially stationary, i.e., it depends only on the distance between positions and time through the spatial distances $h=||s-s^{\prime}||$. Using this structure, we can represent a  spatio-temporal model by the  hierarchical representation given by:

\begin{equation}
\begin{array}{c}
y\left(s,t\right)=z\left(s,t\right)\beta+\xi\left(s,t\right)+\varepsilon\left(s,t\right)\\
\xi\left(s,t\right)=\omega\left(s,t\right)
\end{array}\label{eq:modeloespa=0000E7otemporal-1}
\end{equation}

The $z\left(s,t\right)$ vectors are covariates observed  in locations $\left(s,t\right)$, $\xi\left(s,t\right)$ are the spatial random effects for each $\left(s,t\right)$ and $\varepsilon\left(s,t\right)$  are non-spatial  innovations  with $\varepsilon\left(s,t\right)\sim N(0,\sigma_{\varepsilon}^{2})$. Although this first representation contains only regression effects in the conditional mean, can be extented to include dynamic components, as we will do in this work.
 The $\omega\left(s,t\right)$  process is a random field given by:

\begin{equation}
Cov\left(\omega\left(s,t\right)\left(s^{\prime},t^{\prime}\right)\right)=\left.\left\{ \begin{array}{c}
0\,\,\, if\,\,\,\, t\neq t^{\prime}\\
\sigma_{\omega}^{2}\mathcal{C}\left(h\right)if\,\,\,\, t=t^{\prime}
\end{array}\right\} \right.\label{eq:matern-2}
\end{equation}

\noindent  and  $\mathcal{C}\left(h\right)$ is a covariance function of the Matérn class:

\begin{equation}
\mathcal{C}\left(h\right)=\frac{1}{\Gamma(\nu)2^{\nu-1}}\left(kh\right)^{\nu}K_{\nu}\left(kh\right)
\end{equation}

\noindent  with  $K_{\nu}$ a modified Bessel function of the second type (e.g.  \cite{Abramowitz-Stegun}). The marginal variance is given by: 

\begin{equation}
\sigma_\omega^{2}=\frac{\Gamma\left(\nu\right)}{\Gamma\left(\nu+d/2\right)\left(4\pi\right)^{d/2}\kappa^{2\nu}\tau^{2}}
\end{equation}

\noindent where $\tau$ is a parameter.

The Matérn covariance contains some other spatial covariance functions as particular cases. The  exponential covariance is obtained with $\nu=1/2$,  $d=1$ and $\alpha=1$,
or $d=2$ and $\alpha=3/2$.  There are alternative representations of this covariance function. A first form  is to use a standard deviation parameter $\sigma$ and a
range parameter $\rho$, with $\rho=(8\nu)^{1/2}/\kappa$ being the distance to which the correlation function falls to approximately 0.13, assuming $\nu>1/2$ (e.g.   \cite{Lindgren2015}).  Based on  \cite{spde} we use the following parameterization in terms of $\log\tau$ and $\log\kappa$ for the covariance function:

\[
\log\tau=\frac{1}{2}\log\left(\frac{\Gamma(\nu)}{\Gamma(\alpha)(4\pi)^{d/2}}\right)-\log\sigma-\nu\log\rho
\]

\[
\log\kappa=\frac{\log(8\nu)}{2}-\log\rho
\]

The main advantage of this form is that, conditional on the value of $\nu$,  it results in only two parameters to be estimated.  \cite{Lindgren2015} discuss this property and  the interpretation of parameters in this class of models.   

 This formulation is also interesting because permits the use of Bayesian estimation methods. Stacking the observations of vectors $y\left(s,t\right)$, $z\left(s,t\right)$ and $\xi\left(s,t\right)$  as $\bm{y}$,  $\bm{z}$ and $\bm{\xi}$,  the posterior distribution  of the spatio-temporal model,  in terms of a constant of proportionality, can be written as: 

\[
\pi(\theta,\bm{\xi}|\bm{y})\propto\pi(\bm{y}|\bm{\xi},\theta)\pi(\theta)
\]

Assuming independent prior distributions for $\pi(\theta)$, and exploring the spatial Markov property, the elements of $\bm{y}$  are conditionally independent, and the posterior distribution is given by:

\[
\pi(\theta,\bm{\xi}|\bm{y})\propto\left(\prod_{t=1}^{T}\pi(\bm{y}|\bm{\xi},\theta)\right)\pi(\theta).
\]

Under the  Gaussian Markov random field structure,  this posterior can be represented by:

\[
\pi(\theta,\bm{\xi}|\bm{y})=(\sigma_{\varepsilon}^{2})^{d/2}\exp\left(-\frac{1}{\sigma_{\varepsilon}^{2}}\left(\bm{y}-\bm{z}\beta-\bm{\xi}\right)^{\prime}\left(\bm{y}-\bm{z}\beta-\bm{\xi}\right)\right)
\]

\[
\times\left(\sigma_{\omega}^{2}\right)^{-d/2}|\tilde{\Sigma}|^{-1/2}\exp\left(\frac{1}{2\sigma_{\omega}^{2}}\bm{\xi}^{\prime}\tilde{\Sigma}\bm{\xi}\right)
\]

\[
\times\prod_{i=1}^{dim\left(\theta\right)}\pi\left(\theta_{i}\right)
\]

\noindent with the component $\tilde{\Sigma}$  a $d-$dimensional covariance matrix  with elements $\sigma_{\omega}^{2}\mathcal{C}\left(||h||\right)$.

The continuous spatial component is represented by the computational basis expansion discussed previously, using the finite element representation of the spatial continuum in a triangulation. \cite{spde}) show that the Matérn field $\omega_{t}$ is a Markov random field with a Gaussian $N(0,Q_{S}^{-1})$ distribution, where the matrix $Q_{S}$ is obtained from the solution of the associated stochastic partial differential equation. We first assume that $Q_{S}$ is invariant and has dimension given by the number of vertices of the triangulation, but this structure can be modified to represent time varying precision matrices or spatially non-stationary covariance structures. We discuss a non-stationary spatial formulation in Section \ref{nonstatsec}.

The dynamic formulation used in this paper generalizes the formulation given in Equation (\ref{eq:modeloespa=0000E7otemporal-1})  by including the components of trend, seasonality and cycle in a formulation analogous to the so-called basic structural model of  \cite{Harvey1989}. In this case we generalize the spatial model  by adding the components $\mu_t$, $s_t$ and $c_t$  in the first equation of (\ref{eq:modeloespa=0000E7otemporal-1}), and with dynamics given by the state equations:

\begin{equation}
\begin{array}{c}
\mu_t=\mu_{t-1}+\eta_\mu\\
s_t=s_{t-1}+s_{t-2}+...s_{t-m-1}+\eta_s\\
c_t=\phi_1 c_{t-1}+ \phi_2 c_{t-2}+\eta_c\\
\end{array}\label{state}
\end{equation}

\noindent  as discussed in Section \ref{intro}. The  Bayesian estimation of generalized dynamic models is a well-studied topic in the time-series literature. In particular,  we can estimate the model containing the trend, seasonality and cycle structure and spatial random effects using the method of integrated nested Laplace approximations proposed in  \cite{Rue2009}. The posterior distribution of the complete spatio-temporal model is obtained joining the representation of dynamic models using the INLA formulation of dynamic models proposed by \cite{Rue2010} with the representation spatio-temporal models presented in \cite{Cameletti2013}.  In the following section we show the basic aspects of this method.

\subsection{Bayesian estimation using Integrated Nested Laplace Approximations - INLA\label{INLA}}

The integrated nested Laplace approximations (INLA), introduced by  \cite{Rue2009}, are based on a sequence of Laplace approximations to estimate parameters and latent factors in GMRF structures. This method is very useful since it avoids some problems related with the use of Markov chain Monte Carlo methods, the main tool in the estimation of complex Bayesian models. The INLA is based on deterministic approximations, bypassing the chain convergence problems present in MCMC methods, usually with a much smaller computational cost. The INLA method also resolves some problems of using Laplace approximations, with the use of corrections to the location and skewness problems associated with the naive Gaussian approximation in the Laplace approximation. A very detailed discussion of the INLA method can be found in  \cite{Rue2009}, and a review of the successes and limitations of this method can be found in \cite{Rue2017}. We present only the main steps of the INLA method in this section. Assuming a representation given by:

\begin{equation}
\pi(\xi_{i}|Y)=\int\pi(\xi_{i}|\theta,Y)\pi(\theta,Y)d\theta\label{eq:inla1-1-1}
\end{equation}

\noindent and

\begin{equation}
\pi(\theta_{j}|Y)=\int\pi(\theta|Y)d\theta_{-j}\label{eq:inla2-1-1}
\end{equation}

\noindent  with $\xi$ and  $\theta$ denoting latent factors and hyperparameters, the INLA approximation  consists of a  sequence of analytical  Laplace approximations  or the conditional distributions $\pi(\theta|Y)$ and $\pi(\xi_{i}|Y)$. The first step is to approximate the  conditional distribution of latent factors  $\pi(\xi|Y,\theta)$ with a multivariate Gaussian distribution  $\hat{\pi}_{G}(\xi|Y,\theta)$, evaluated in the mode. Based on this approximation,  the posterior distribution  of  $\theta$   is estimated using another Laplace approximation: 

\begin{equation}
\tilde{\pi(}\theta|Y)\propto\frac{\pi(\xi,\theta,Y)}{\tilde{\pi}_{G}(\xi|\theta,Y)}\mid_{\xi=\xi^{*}\left(\theta\right)},\label{eq:inla3-1-1}
\end{equation}

\noindent with $\xi^{*}\left(\theta\right)$  indicating that again this approximation is made in the mode of the conditional distribution of $\xi|\theta$. This mode is evaluated using numerical   Newton-Raphson root searching algorithms. Using these initial results, the Laplace approximation is now applied to the conditionals  $\pi(\xi_{i}|\theta,Y)$, for a sequence of $\theta$ values. The posterior distributions of the latent factors are calculated in similar form by:

\begin{equation}
\tilde{\pi}_{LA}(\xi_{i}|\theta,Y)\propto\frac{\pi(\xi,\theta,Y)}{\tilde{\pi}_{G}(\xi_{-i}|\xi_{i},\theta,Y)}\mid_{\xi_{-i}=\xi_{-i}^{*}\left(\xi_{i},\theta\right)},\label{eq:inla4-1-1}
\end{equation}

\noindent  where $\xi_{-i}$  denotes the latent factors $\xi$ with the $i$-th element omitted,  and $\tilde{\pi}_{G}(\xi_{-i}|\xi_{i},\theta,Y)$ a Gaussian approximation of  $\pi(\xi_{-i}|\xi_{i},\theta,Y)$ maintaining $\xi_{i}$  and $\xi_{-i}^{*}\left(\xi_{i},\theta\right)$ at the mode of  $\pi(\xi_{-i}|\xi_{i},\theta,Y)$, and the final step is the combination of the previous approximations  and the integration of irrelevant factors. The marginal posterior distribution of latent factors is obtained as:

\[
\pi(\xi_{i}|Y)=\int\pi(\xi_{i}|\theta,Y)\pi(\theta,Y)d\theta\thickapprox\sum\pi(\xi_{i}|\theta_{k},Y)\tilde{\pi(}\theta_{k}|Y)\triangle_{k}
\]

\noindent with $\triangle_{k}$ denoting the grid values used in the numerical approximation. Analogously, the marginal posterior distribution of the hyperparameters is estimated by:

\[
\pi(\theta_{j}|Y)=\int\pi(\theta|Y)d\theta_{-i}\thickapprox\int\tilde{\pi(}\theta_{k}|Y)d\theta_{-i}.
\]

The estimation of generalized dynamic models and basic structural models using the INLA method is possible due to the additive nature of these components. Assuming that these components are independent, the estimate is given by the inclusion of these components in the likelihood function of the process and in the structure of priors. A detailed discussion of the estimation of dynamic models using INLA can be seen in \cite{Rue2010}. As discussed in the previous section, Bayesian estimation of the continuous spatial model is possible by the representation of the spatial random effects through the computational representation of the GMRF through the $Q$ precision matrix using the finite element structure in a triangulation, as discussed in \cite{spde}. This work combines these two elements to represent continuous spatial dynamic processes using a spatial generalization of the basic structural model of  \cite{Harvey1989}.

In all estimation procedures,  we use a set of independent priors, with Gaussian distributions for the parameters of the conditional mean and autoregressive parameters, log-gamma distributions for the parameters of the spatial covariance function, and gamma distributions for the precision parameters. The hyperparameters in these priors are available from us, and are based on the default values for the spde models used in the R-INLA implementation of the spde method of  \cite{spde}. In general,  the results are robust to the choice of hyperparameters in the prior distributions.

\section{Dataset\label{dataset}}

We use in this work the Brazilian meteorological data provided by the National Institute of Meteorology (INMET), through the BDMEP system, available at \href{ http://www.inmet.gov.br/portal/index.php?r=bdmep/bdmep}{ http://www.inmet.gov.br}. This system provides daily and monthly digital meteorological data of historical series from the various conventional meteorological stations of the INMET station network, with measurements in accordance with the international technical standards of the World Meteorological Organization. In the BDMEP system, daily and monthly data from 1961 on are available on precipitation occurred in the last 24 hours; dry bulb temperature; wet bulb temperature; mean and maximum temperature; relative humidity; atmospheric pressure at station level; insolation; and wind direction and speed.  We use this database to avoid the use of fixed grid data, as is usual in climate data sources, as the climatic data provided by the GISS Surface Temperature Analysis \href{(https://data.giss.nasa.gov/gistemp/}{(https://data.giss.nasa.gov/gistemp/}). Gridded data involve interpolation procedures and artificially increases the accuracy of the data, and our goal is to capture the information actually observed by monitoring stations.

In this work, we present the estimation results for the Northeast region of Brazil for the series of average temperature, maximum temperature and rainfall, whose calculation method is described in the following \href{http://www.inmet.gov.br/webcdp/climatologia/normais/imagens/normais/textos/metodologia.pdf}{link}. Although the data are available in daily and monthly frequencies, we use a quarterly aggregation of the monthly data in this work, to facilitate visualization of the results. We analyze the mean of temperatures observed in each quarter and the rainfall accumulated throughout the quarter in the analysis. Due to space constraints, we only present the results in quarterly frequency for the Northeast region of Brazil. The results for the other regions of Brazil and monthly frequency are available on request.

\begin{figure}
\caption{ \textbf{Territorial division - Northeast Region - Brazil}\label{mapaestados}}
\begin{center}
\includegraphics[width=120mm]{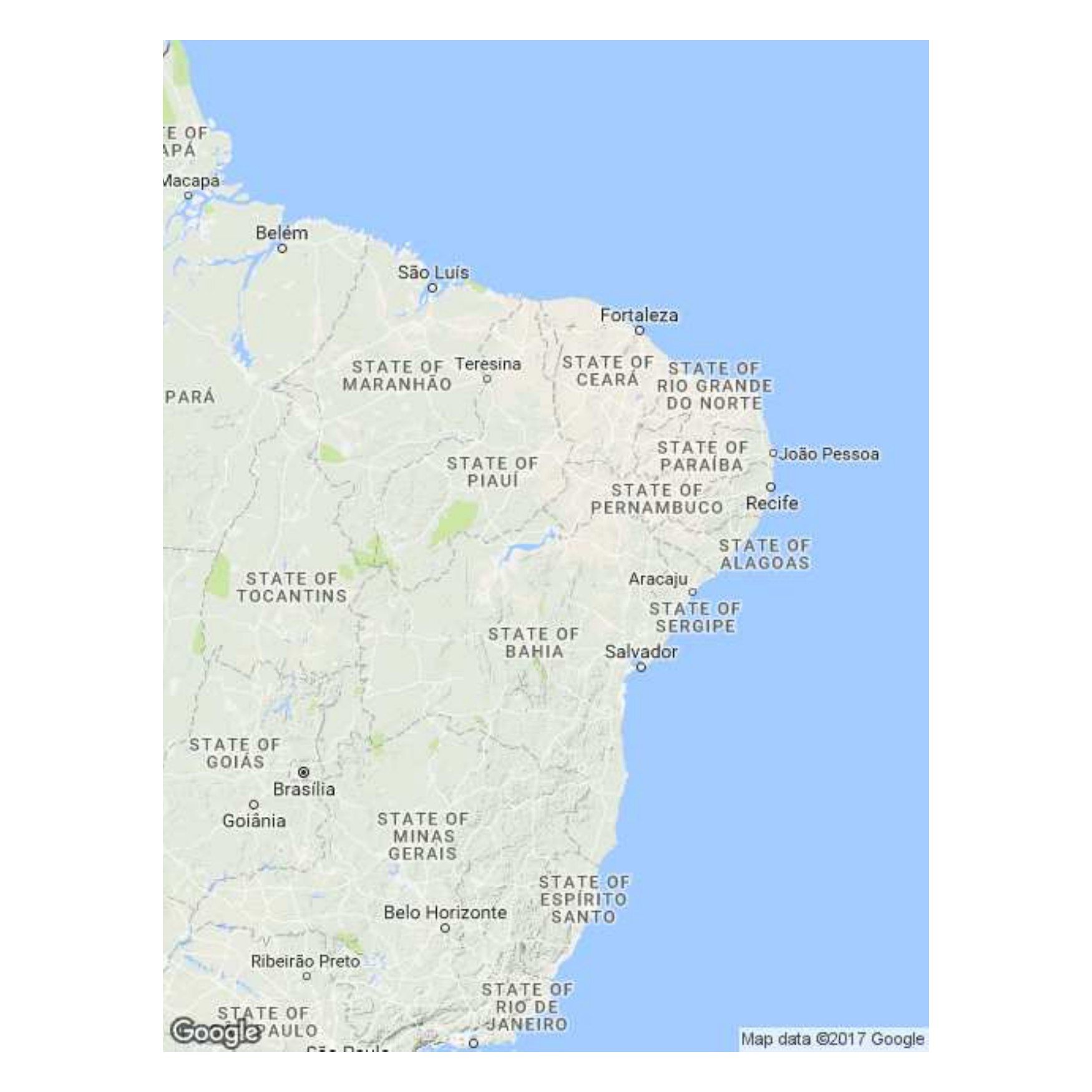}\\
\vspace*{0.0cm}
\begin{footnotesize}
Source: Google Maps
\end{footnotesize}
\end{center}
\end{figure}

\begin{figure}
\caption{ \textbf{Northeast Region -  K{\"o}ppen climate classification}\label{koppen}}
\begin{center}
\includegraphics[width=110mm,angle=270]{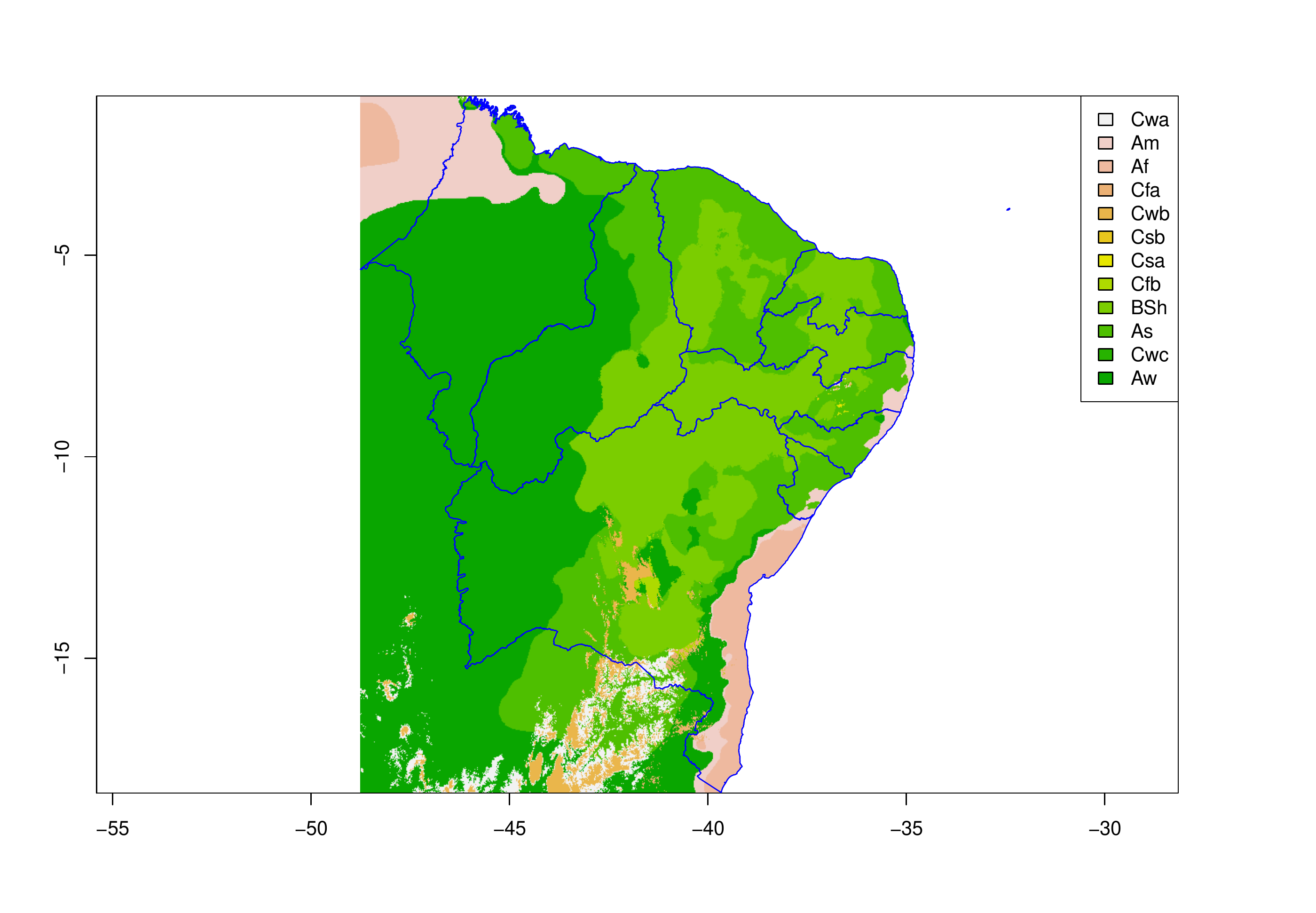}

\begin{footnotesize}
\singlespacing
\textbf{Koppen Climate Classification - Northeast Region in Brazil}\\
Cwa: (C) Humid subtropical  (w) With dry winter (a) and hot summer\\
Am:  (A) Tropical (m) monsoon\\
Af:  (A) Tropical(f) without dry season\\
Cfa: (C) Humid subtropical (f) Oceanic climate, without dry season  (a) and hot summer\\
Cwb: (C) Humid subtropical  (w) With dry winter  (b) and temperate summer\\
Csb: (C) Humid subtropical (s) With dry summer (b) and temperate summer\\
Csa: (C) Humid subtropical (s) With dry summer (a) and hot summer\\
Cfb: (C) Humid subtropical (f) Oceanic climate, without dry season (b) and temperate summer\\
BSh: (B) Dry (S) Semi-arid (h) low latitude and altitude\\
As:  (A) Tropical (s) with dry summer\\
Cwc: (c) Humid subtropical  (w) With dry	winter and (c) short and cool summer\\
Aw:  (A) Tropical (w) with dry winter\\
Source -  \cite{Stape2013}

\end{footnotesize}
\end{center}
\vspace*{0.0cm}
\end{figure}

The Northeast region comprises the states of Alagoas, Bahia, Ceará, Maranhão, Paraíba, Pernambuco, Piauí, Rio Grande do Norte and Sergipe. The territorial division of these states can be seen in Figure \ref{mapaestados}. This region is especially interesting for analysis of climate changes due to the extreme variability of climates observed in the region. The analysis of climate change in this region is also socially important due to the sensitivity of economic activity to climatic variations, presenting repeated periods of intense droughts with serious consequences on farms and water supply. The analyzed region is situated between latitudes (-18.34874, -1.045085) and longitudes (-48.75471, -32.39088), and presents three main climate types - humid coastal climate (coast of Bahia to Rio Grande do Norte), tropical climate (parts of Bahia, Ceará, Maranhão and Piauí), and semiarid tropical climate (Northeast \textit{Sert\~ao}). A complete classification based on the  K{\"o}ppen framework, as constructed by \cite{Stape2013}, is shown in Figure \ref{koppen}. In this classification, we have the various sub-types of climate in this region, with remarkable thermal and precipitation amplitude.

This region presents great inter-annual variability, with extremely dry and rainy years, as well as high inter-seasonal variability. Historical average temperatures range from $20\degree C$ to $28\degree C$, but in the higher regions on the Chapada Diamantina and Borborema Plateau, the average temperatures are below $20\degree C$. The main factors that determine the climate variations are the South Atlantic and North Atlantic subtropical anticyclones and the low pressure equatorial region (\textit{cavado equatorial}), as discussed in  \cite{Molion2002} and  \cite{Iracema2009}. The low frequency variations are related to global atmospheric circulation patterns associated with the El Niño and La Niña phenomena of surface water temperature in the Central and East Equatorial Pacific. In particular, El Niño is associated with the periods with inverse pressure variations at sea level in the Eastern Tropical Pacific. A detailed discussion of the climate of the Northeast region and of Brazil in general can be found in \cite{Iracema2009}. 

\begin{figure}
\caption{ \textbf{Northeast Region - Location of weather stations}\label{stations}}
\begin{center}
\includegraphics[width=90mm,angle=270]{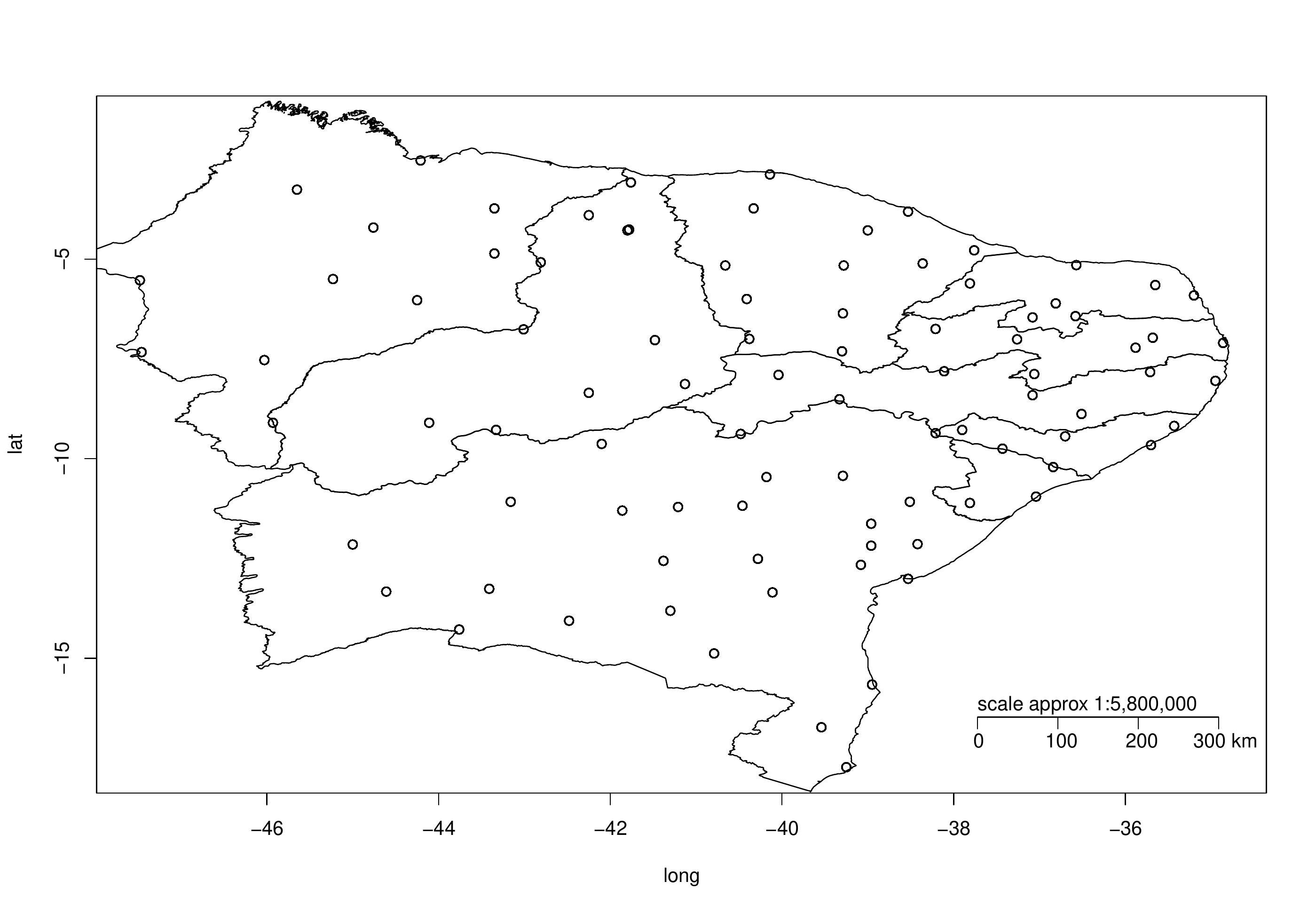}
\vspace*{0.0cm}
\end{center}
\end{figure}

\begin{figure}
\begin{center}
\caption{ \textbf{Proportion of stations with non-null climate observations}\label{proportion}}
\includegraphics[width=0mm,angle=270]{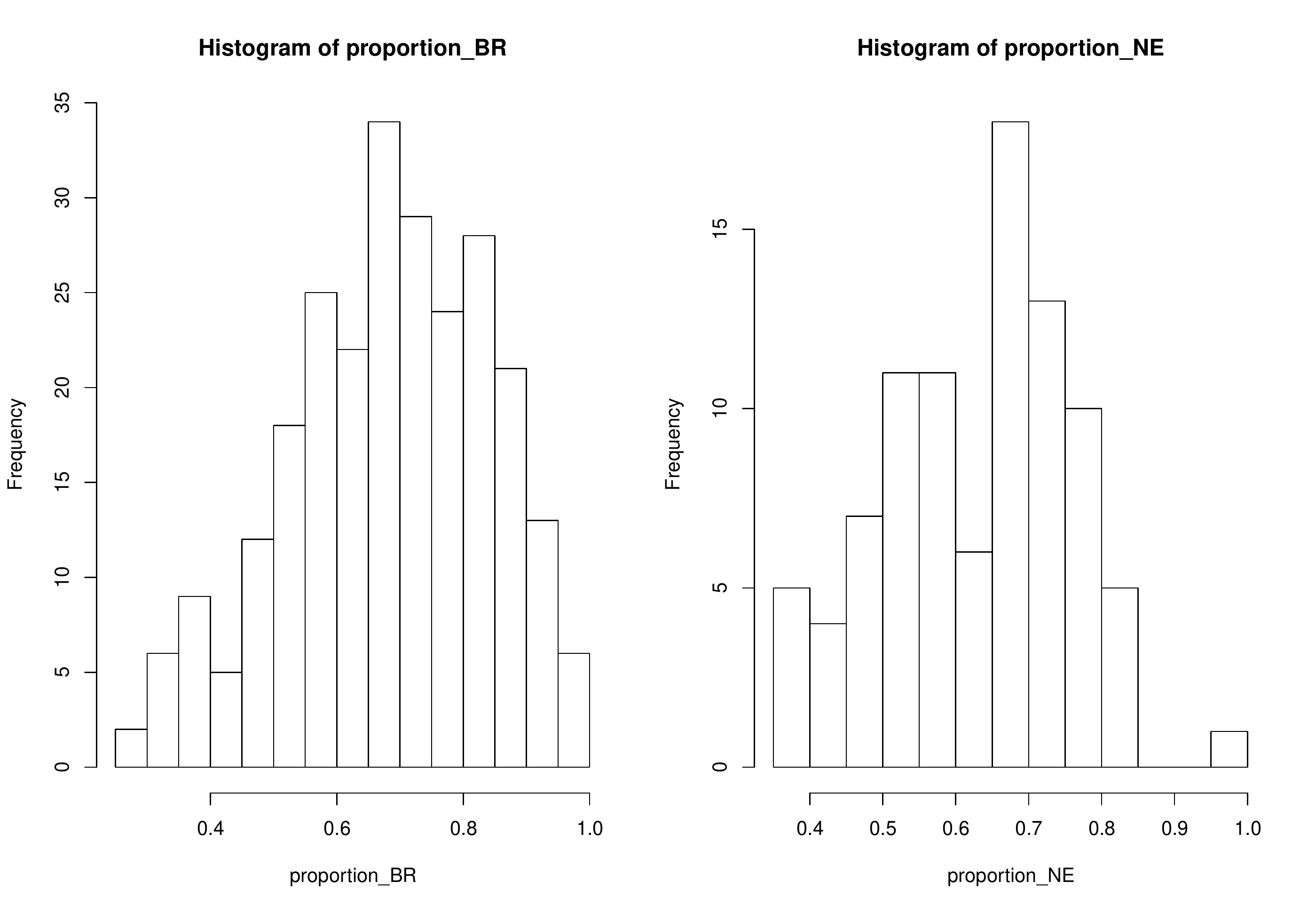}
\includegraphics[width=90mm,angle=270]{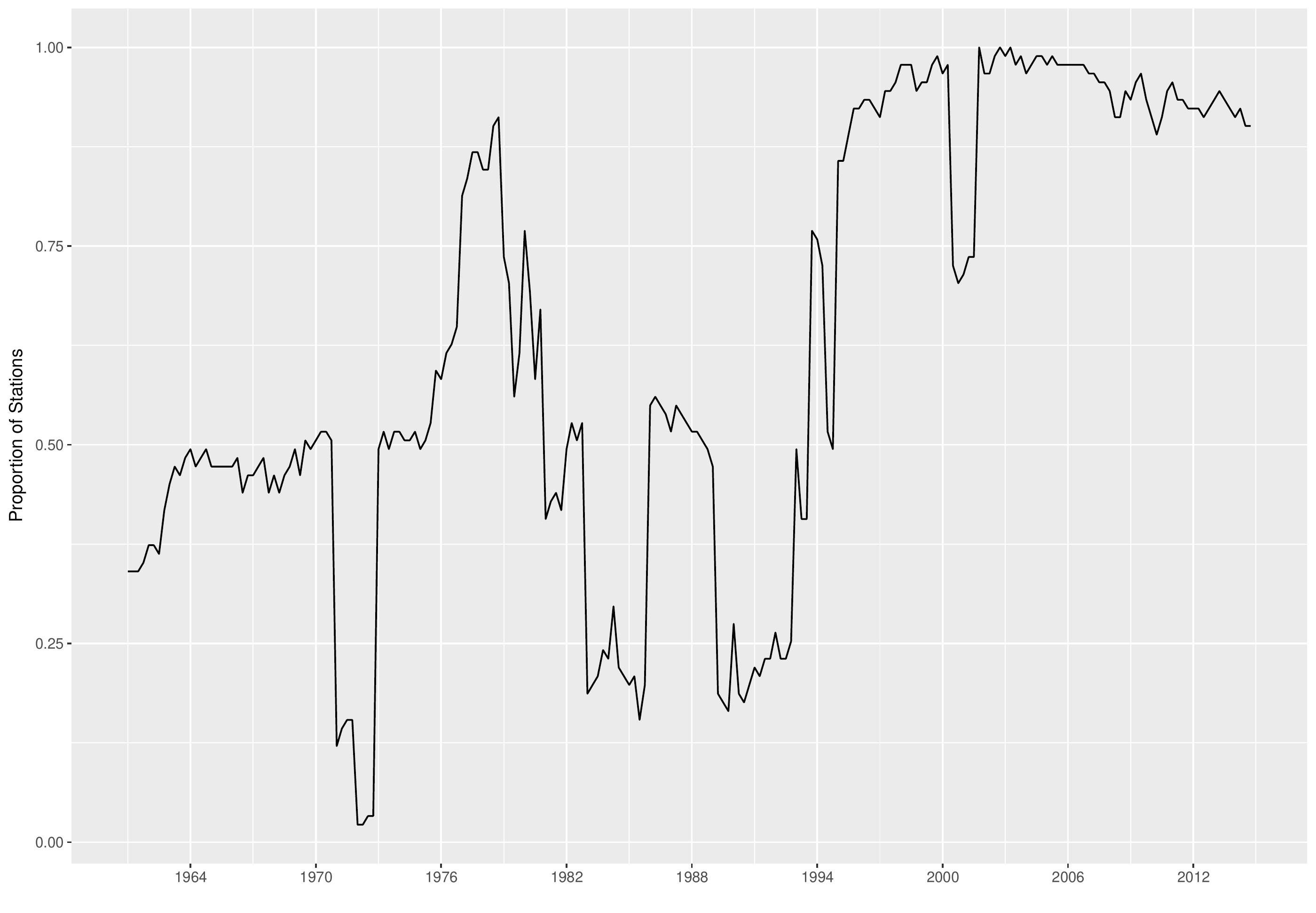}
\vspace*{0.0cm}
\end{center}
\end{figure}

We use data from the 91 weather monitoring stations of the National Institute of Meteorology located in the Northeast region. Figure \ref{stations} shows the distribution of these stations in the analyzed region. As discussed in the introduction, the data provided by these stations are subject to missing data problems, especially for a part of the 1970s. Figure \ref{proportion} shows the proportion of stations with observed data in each period of time. These figures show that this missing data characteristic is a fundamental aspect of this problem. The usual treatments are methods of interpolation or imputation of data, or withdrawing from the sample the stations with a high proportion of missing observations. Note that these treatments can have statistical consequences. Station withdrawal decreases the set of information available in each period, and ad hoc interpolation methods can alter the structure of the process, for example by reducing the observed variability of the data.

\begin{figure}
\caption{ \textbf{Boxplot - Average temperature by year and quarter}\label{boxp1}}
\begin{center}
\includegraphics[width=90mm,angle=270]{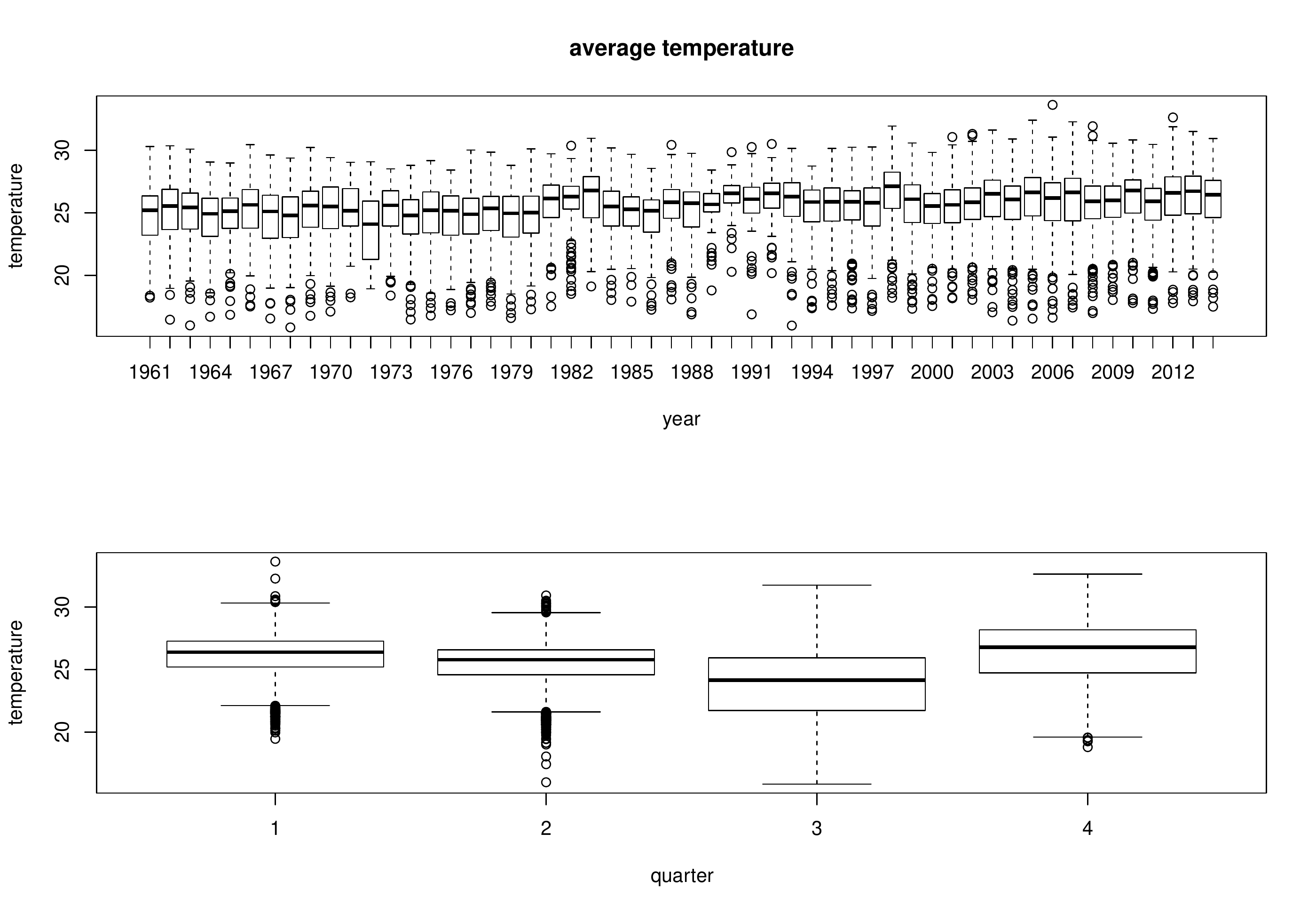}
\vspace*{0.0cm}
\end{center}
\end{figure}

\begin{figure}
\caption{ \textbf{Boxplot -  Maximum temperature  by year and quarter }\label{boxp2}}
\begin{center}
\includegraphics[width=90mm,angle=270]{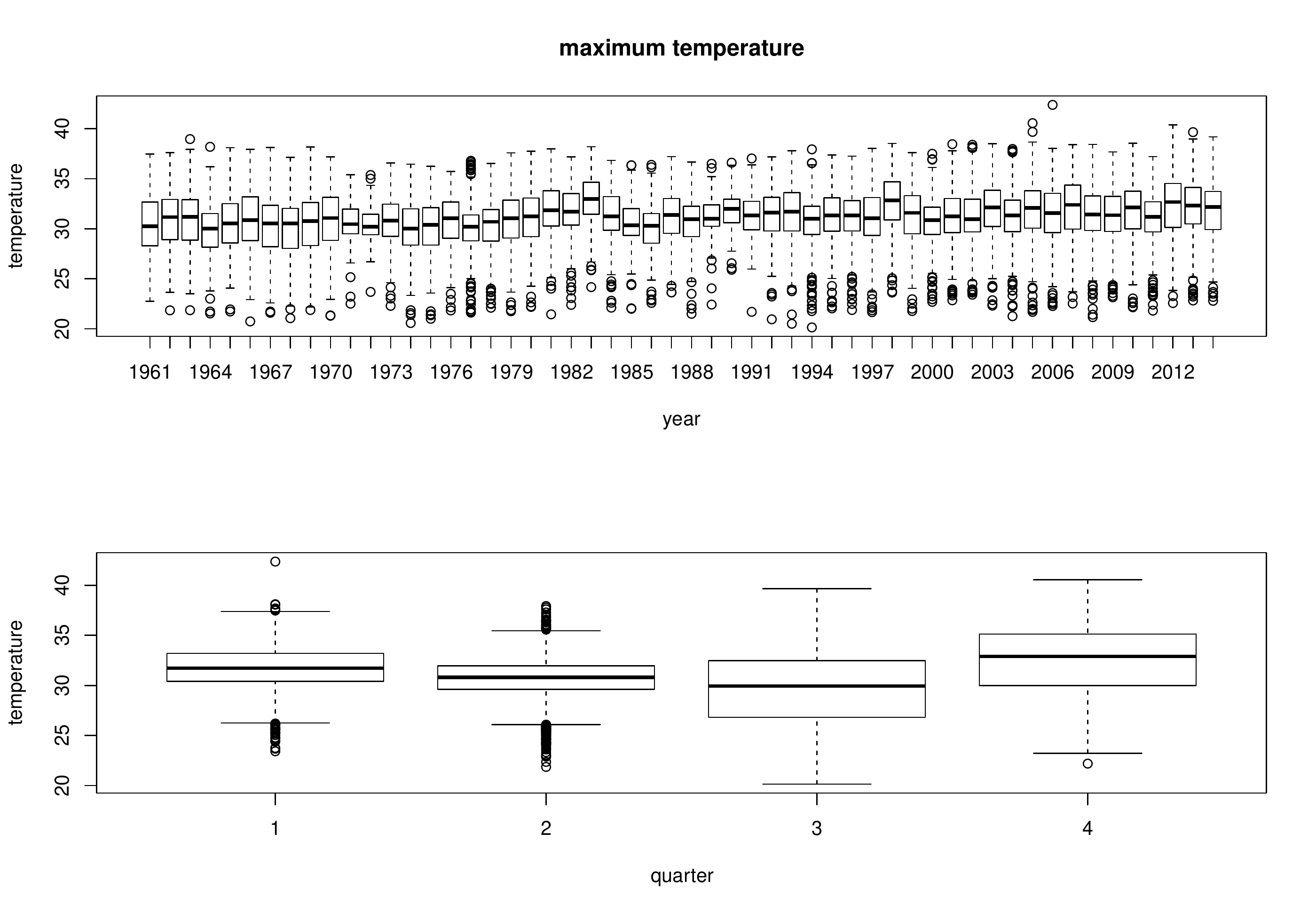}
\vspace*{0.0cm}
\end{center}
\end{figure}

\begin{figure}
\caption{ \textbf{Boxplot - Rainfall by year and quarter}\label{boxp3}}
\begin{center}
\includegraphics[width=90mm,angle=270]{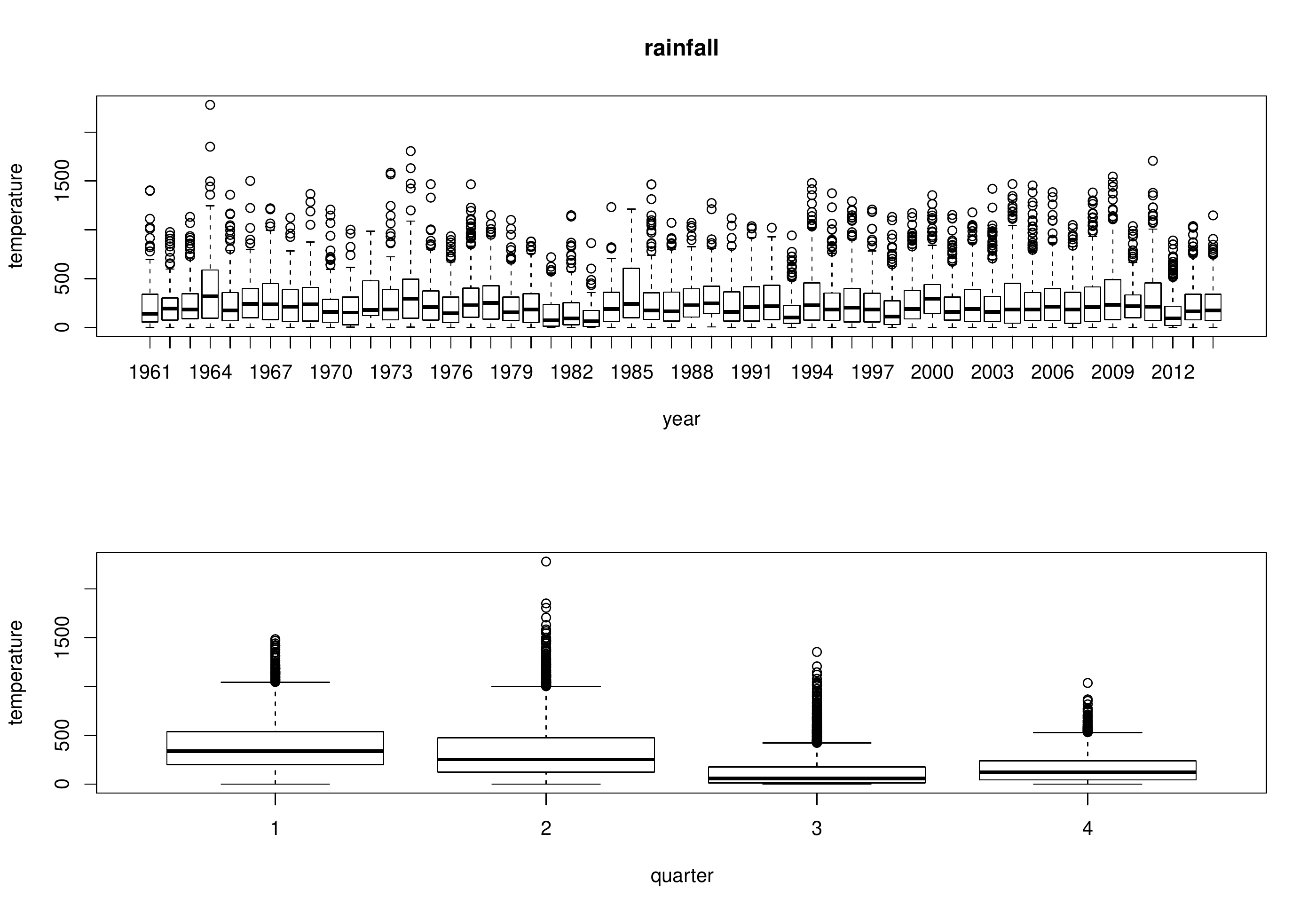}
\vspace*{0.0cm}
\end{center}
\end{figure}

\begin{figure}
\caption{ \textbf{Time series of average and maximum temperature, aggregated by period\label{meantemp}}}
 
\begin{center}
\includegraphics[width=90mm,angle=270]{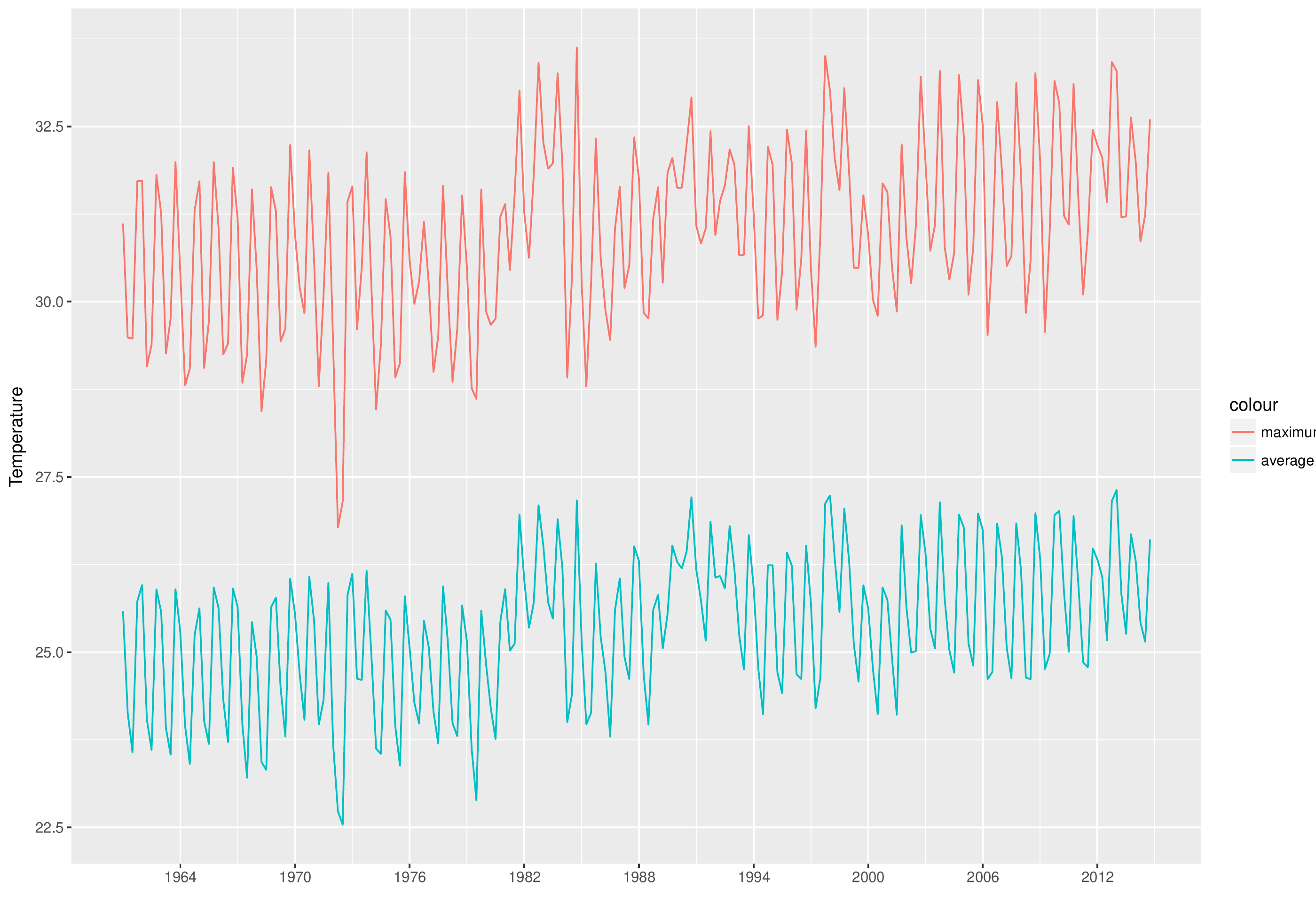}
\vspace*{0.0cm}
\end{center}
\end{figure}

\begin{figure}
\caption{ \textbf{Time series of rainfall,  aggregated by period }\label{meanpluv}}
\begin{center}
\includegraphics[width=83mm,angle=270]{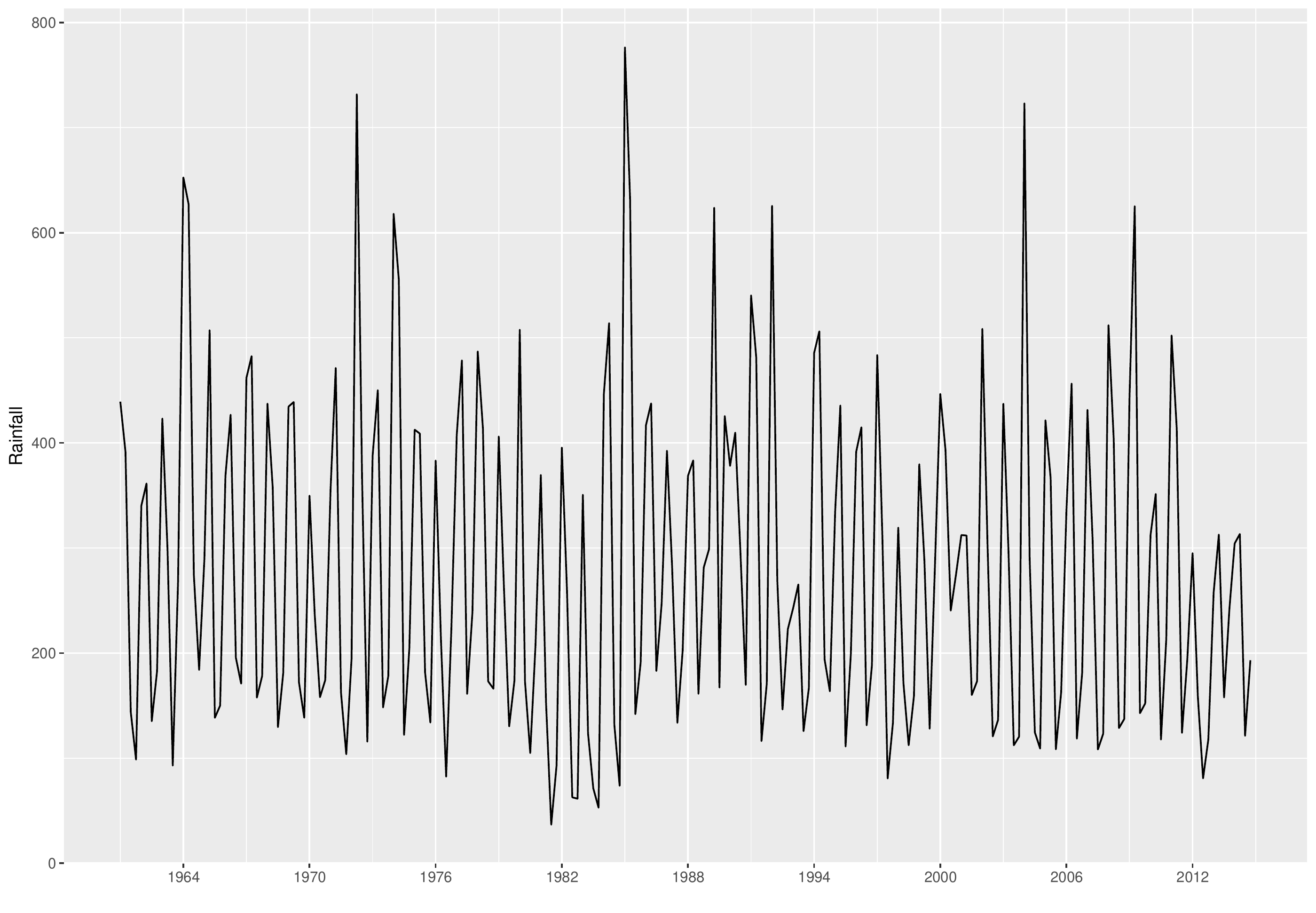}
\vspace*{0.0cm}
\end{center}
\end{figure}

Figures \ref{boxp1}, \ref{boxp2} and  \ref{boxp3} present boxplots representing the distribution of the temperature and rainfall precipitation series by year and also by quarter. In these figures the great variability between years and quarters in these data is evident, showing the large climatic heterogeneity that exists in this process. For ease of visualization, we show in Figures \ref{meantemp} and \ref{meanpluv} an aggregation with the mean values between all the stations for each quarter in the sample. One can observe a possible change in the average and maximum temperature patterns from the 1980s, while it is not possible to observe any clear pattern of change in the rainfall series in this period. This series shows the great variability in rainfall patterns in this region, with periods of severe drought in 1981-1983 and periods with relatively high rainfall. 

\section{Results\label{results}}

\begin{figure}
\caption{ \textbf{Triangulated mesh}\label{mesh}}
\includegraphics[width=100mm,angle=270]{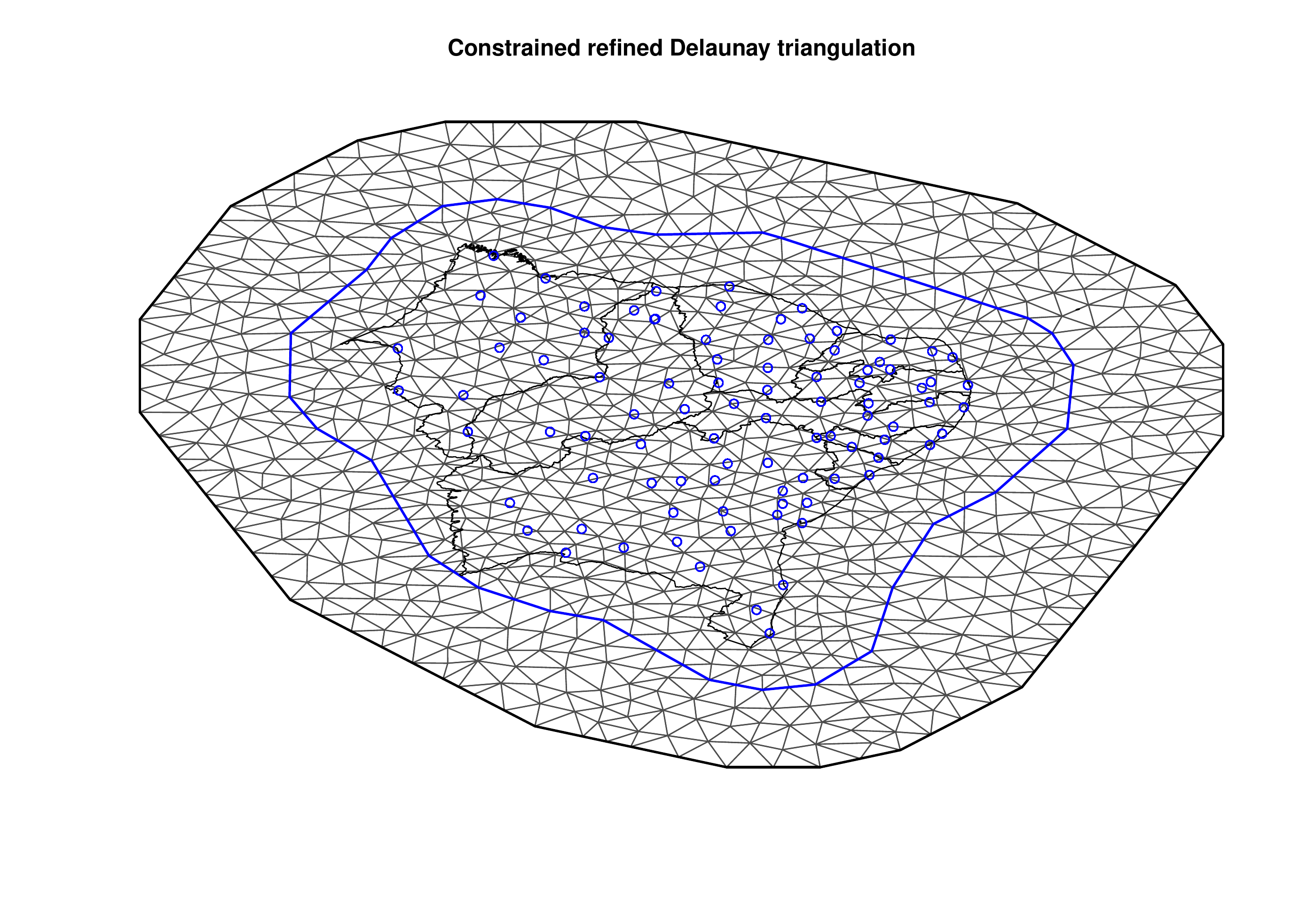}
\vspace*{0.0cm}
\end{figure}

To perform the inference procedures using the method proposed in Section \ref{method}, the first step is to construct a triangulated mesh for the basis representation related to the spde solution. The triangulated mesh used in this work is presented in Figure \ref{mesh}, and represents an approximation of the Northeast region space using 1990 triangles. Note that the mesh also requires an external area, which is necessary to avoid edge problems in numerical approximations, as discussed in \cite{spde}. The mesh construction involves an optimal triangle size that considers the number and distribution of the observed points, and the computational cost involved in the approximation. The chosen size allows an adequate approximation of the continuous spatial process. We try other specifications and the results are robust to the choice of mesh.

In line with other similar studies (e.g. \cite{Stape2013b}), we test a set of explanatory variables in the definition of the best models for temperature and precipitation. The set of variables tested includes altitude, latitude, longitude and distance to the sea for each observation in the sample. This is a common set of variables used in temperature and precipitation modeling, since it allows controlling the main fixed effects related to climate determinants, such as the influence of sea temperature on air temperature and precipitation (e.g.  \cite{Hillebrand2013})

\begin{figure}
\caption{ \textbf{Altitude (elevation)  map - Northeast Regions}\label{altimetria}}
\begin{center}
\includegraphics[width=120mm,angle=270]{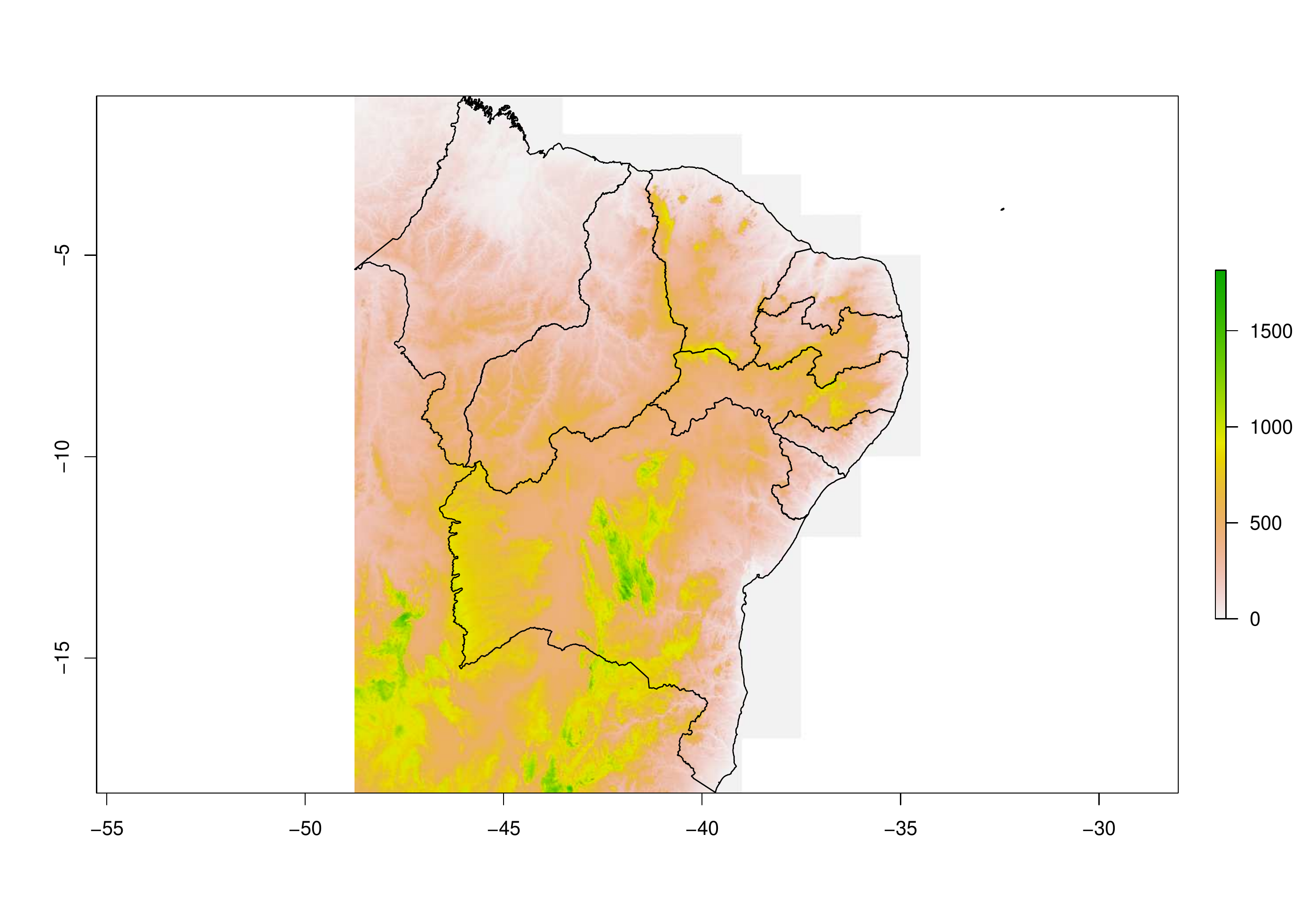}\\
\vspace*{0.0cm}
\begin{footnotesize}
 Source - National Institute for Space Research (INPE) - Digital Elevation Model (DEM), TOPODATA Project - \href{http://www.dsr.inpe.br/topodata/}{http://www.dsr.inpe.br/topodata/}.  Altitude measured in meters. 
\end{footnotesize}
\end{center}
\end{figure}

We use mainly the deviation information criteria - DIC ( \cite{Spiegelhalter2002}) to perform model selection. According to this criterion, the model selected for the series of average and maximum temperatures contains altitude, latitude and distance to the sea as explanatory variables, while for rainfall  the DIC  of the chosen model includes only altitude. Figure \ref{altimetria} shows the high resolution digital topography map used in this work, provided by the National Institute for Space Research (INPE), \href{http://www.inpe.br/}{http://www.inpe.br/}. Our method allows constructing projections for the variable of interest in each period and spatial location, for which it is necessary to measure the value of the explanatory variables at each point in the continuum. 

The specification of the structural decomposition of time series is based on an additive process of trend, seasonality and cycle, analogous to the so-called basic structural model (e.g., \cite{Harvey1989} and \cite{Proietti1991}), following the general specification given by Equation (\ref{model1}). The trend series is modeled as a first-order random walk process, also known as the local-level model. This specification is widely used in modeling climatic processes, as discussed in \cite{Gordon1991},  \cite{Hillebrand2013} and  \cite{Proietti2017}, and is the main component for interpreting changes in the permanent patterns of the climatic series in this study. The seasonality process is based on a formulation of mean effects by period, with the restriction that these effects most sum to zero. Another possibility would be to use a structure based on trigonometric decompositions, as used for example in \cite{Proietti2017}. To represent a possible cyclic component, we use a formulation similar to the so-called unobserved component models (\cite{Clark1987}), where the cyclic component is represented by a latent factor with a second-order autoregressive (AR) structure. The AR(2) formulation allows capturing cyclic patterns if the roots of the lag operator's polynomial form are complex, which is equivalent to periodic patterns in the dependence structure. For computational reasons we represent the AR(2) process through its partial autocorrelations functions. The spatial covariance structure is modeled by the continuous Matérn covariance function, parameterized by two parameters, as discussed in Section \ref{method}.

In this model, the parameters to be estimated are the $\beta$ parameters associated with the explanatory variables, the precision of the non-spatial error components (Precision Gaussian), the precision of the trend components (Precision RW), seasonality (Precision Seasonal) and cycle (Precision Cycle), corresponding to the inverse of variance of innovation components $\epsilon\left(s,t\right)$, $\eta_\mu$, $\eta_s$ and $\eta_c$ in Eq. (\ref{model1}). The parameters of the second-order autoregressive process of the cycle are parameterized as partial autocorrelations (PACF1 and PACF2), and the parameters of spatial covariance are represented by log $\tau$ and log $\kappa$, with the use of log transformation to ensure positivity in the estimated parameters.

\begin{table}
\protect\caption{\label{tab:mean} \textbf{Estimated parameters - Average temperature} }
\begin{small}
\begin{tabular}{ccccccc}
\hline 
 & {\small{}mean} & {\small{}sd} & {\small{}.025q} & {\small{}.5q} & {\small{}.975q} & {\small{}mode}\tabularnewline
\hline 
{\small{}Altitude} &  -0.0076 & 0.0002 &   -0.0081 & -0.0076 &   -0.0072& -0.0076 \tabularnewline
{\small{}Latitude } & 0.1287& 0.0640 &    0.0031&   0.1287 &    0.2542&  0.1287 \tabularnewline
{\small{}Distance to Sea } & 0.0044 &0.0010  &   0.0024&   0.0044&     0.0064 & 0.0044    \tabularnewline
Precision Gaussian &   0.9655& 1.270e-02 &    0.9398 &    0.9659 &    0.9895  &  0.9671\tabularnewline
Precision RW &979.8983& 7.055e+02 &  162.8741 &  811.2303&  2791.3847 & 458.2373\tabularnewline
Precision Seasonal  &  29883.5201 &2.406e+04  &3089.2646& 23719.2389& 91663.4127& 9086.3580\tabularnewline
Precision Cycle  &5.3213 &8.881e-01 &    3.9058 &    5.2027 &    7.3699 &   4.9430\tabularnewline
PACF1  & 0.2891& 2.269e-01 &   -0.2646 &    0.3409   &  0.5797 &   0.5068\tabularnewline
PACF2 &  -0.0460 &7.860e-02 &   -0.1920 &   -0.0493 &    0.1160&   -0.0593\tabularnewline
log $\tau$ &  -1.0090 &1.378e-01 &   -1.3168 &   -0.9918 &   -0.7846 &  -0.9395\tabularnewline
log $\kappa$ & -0.1623 &2.686e-01&    -0.6009&    -0.1958 &    0.4381&   -0.2937\tabularnewline
Marginal Lik. &-18170.40    & obs  & 12323 &  &  & \tabularnewline
DIC & 35676.84 &  &  &  &  & \tabularnewline
\hline 
\end{tabular}
\end{small}
\begin{tiny}\\
\end{tiny}
\end{table}

\subsection{Results - Average Temperature}

The results of the estimation for the average temperature data are shown in Table \ref{tab:mean}, and are based on 12323 observations for the period 1961-2014. The estimated parameters indicate a negative relation between temperature and altitude, as expected, and also a positive relation between latitude and temperature. As the entire analyzed region is below the equator, the latitudes are negative, and in locations more to the south, the temperature is lower, again an expected result. The distance to the sea is estimated with a positive parameter, which can be explained by the effect of sea breezes on the average temperature; with the longest distance, this effect is smoothed.

\begin{figure}
\caption{ \textbf{Trend, seasonal and cycles decomposition - Average temperature}\label{bsmmean}}
\begin{center}
\subfloat[Trend]{\includegraphics[width=80mm]{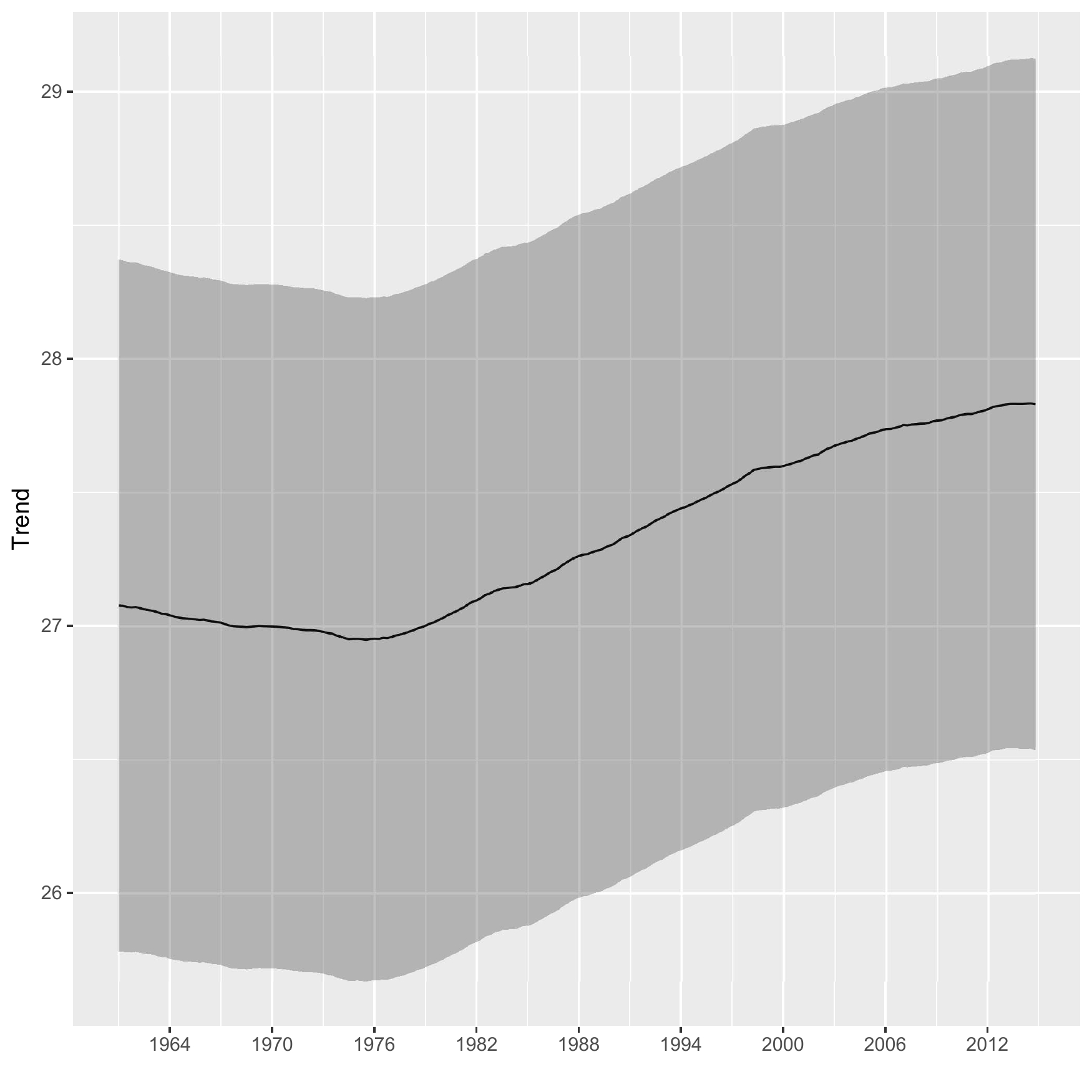}}\subfloat[Seasonal]{\includegraphics[width=80mm]{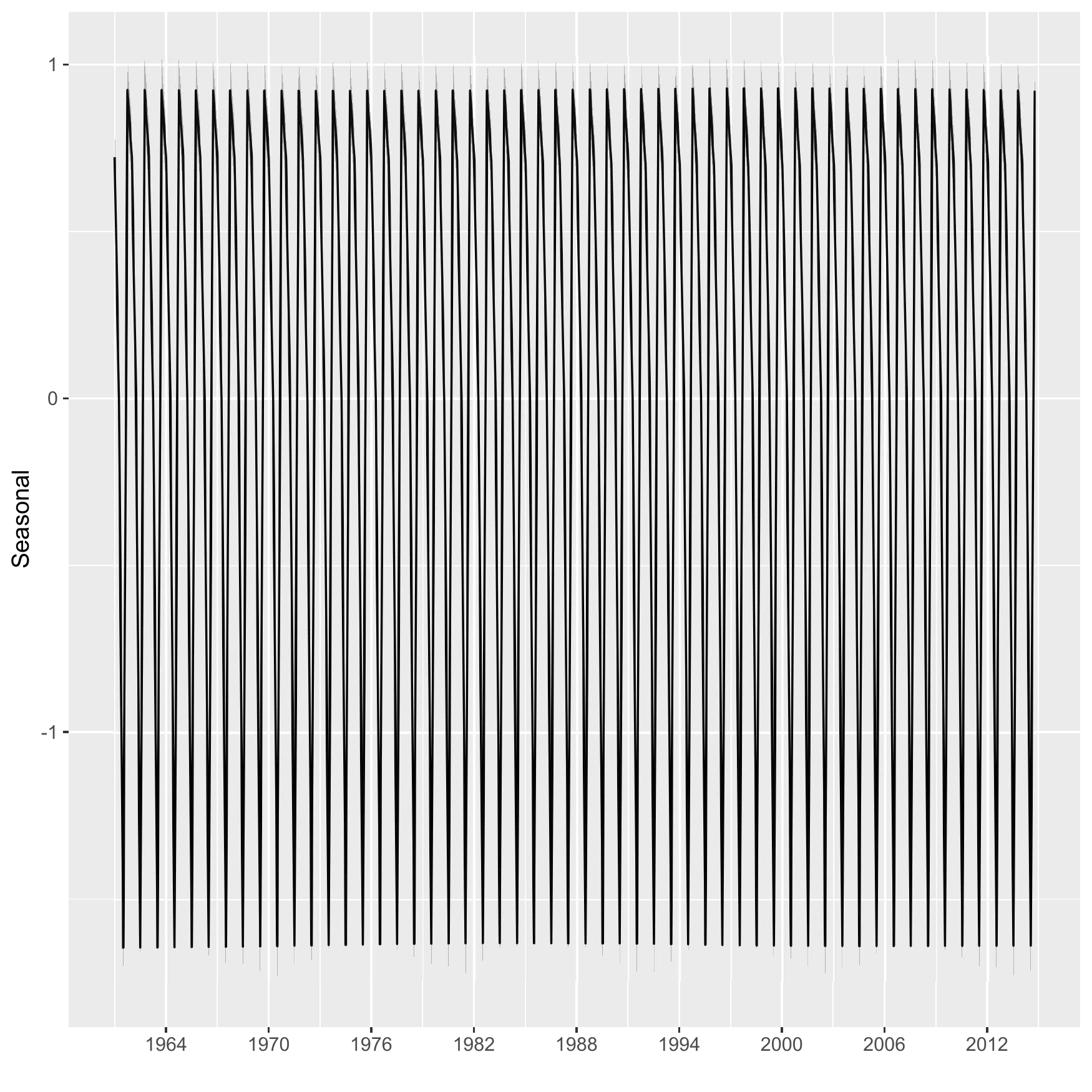}}\\
\subfloat[Cycle]{\includegraphics[width=80mm]{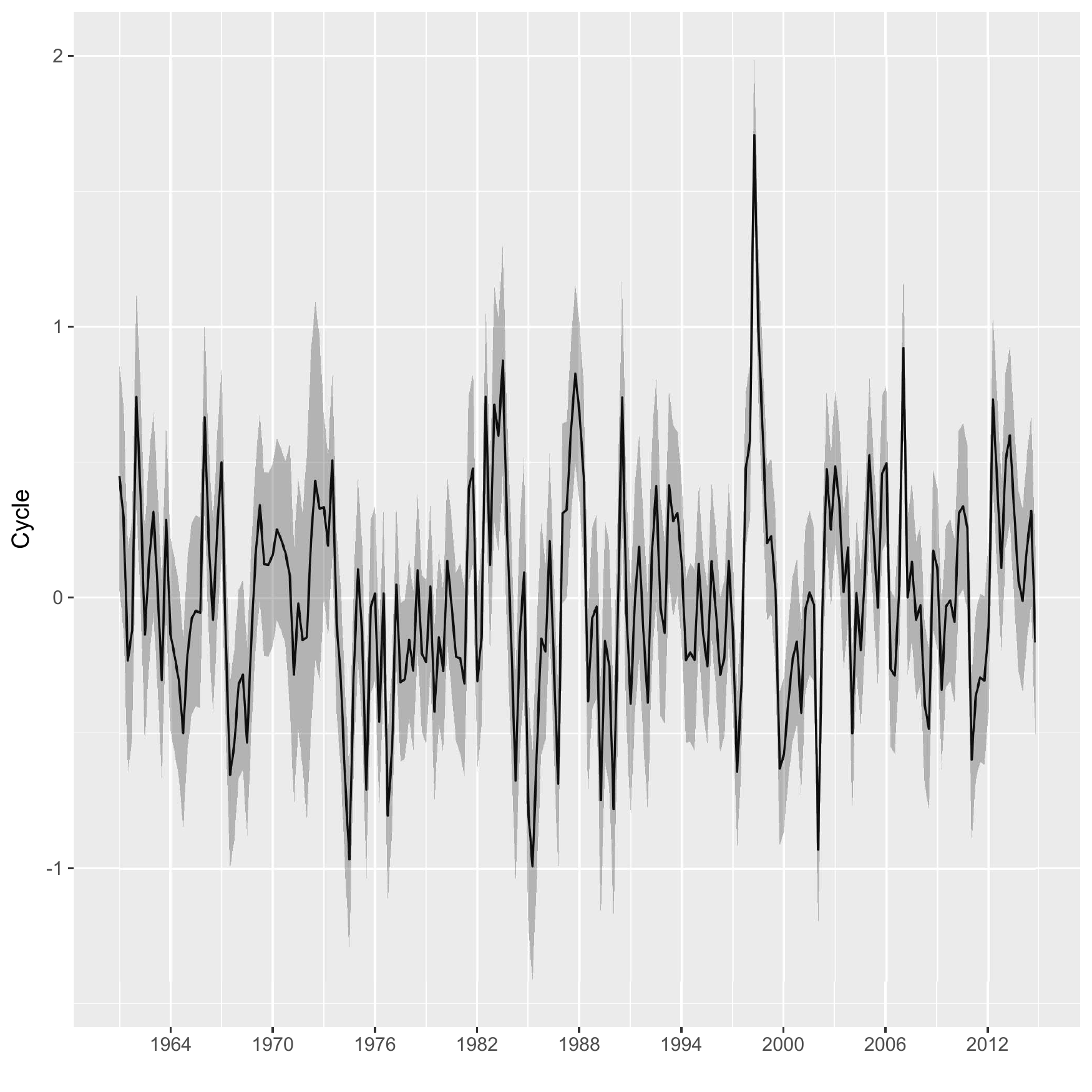}}
\vspace*{0.0cm}
\end{center}
\end{figure}

The estimated precision parameters indicate high precision for trend and seasonality components, and relatively minor precision for the components of cycle and non-spatial (Gaussian) error. The interpretation of these results is most convenient through the estimated trend, seasonality and cycle components, shown in Figure \ref{bsmmean}, which presents the posterior mean of the estimated components and the associated 95\% Bayesian credibility interval.

\textbf{Note} - An important point is about the interpretation of the estimated values for the trend. The trend is estimated conditional to the effect of the explanatory variables. Thus, to obtain an interpretation of trend as mean temperature, one must adjust for these effects, for example by calculating the effect of the explanatory variables on the average of the observed values in each explanatory variable. The average altitude for the stations is 297.37 meters, the average latitude is -8.40 and the average distance to the sea of 213.60km. Thus,  the correction to obtain the adjusted trend for the covariates in these averages would be $-0.0076 * 297.37 + 0.1287 * -8.40 + 0.0044 * 213.60 = -2.4012$. With this adjustement it's possible to compare the fitted trend with the mean values of temperature, for example the values show in Figures \ref{boxp1} and \ref{meantemp}.

The most notable result is the growth pattern observed in the trend from the mid-1970s. Since this component represents the permanent pattern of temperature change (e.g.,  \cite{Gordon1991}), the model captures growth of about $0.8\degree C$ in the average temperatures in about 30 years, supporting the evidence of global warming observed in this period, and supporting the presence of climate change. This acceleration pattern observed in the temperature growth in the 1980s is consistent with the results found in \cite{Ji2014}, supporting the particular warming patterns found in semi-arid regions shown in this work. The value of increase in temperature found is also compatible with other studies of global climate change, e.g., \cite{Estrada2013}. Note that the credibility intervals capture the uncertainty associated with a nonstationary component for the trend, and this range is compatible with the heterogeneity observed in the temperature series across time and monitoring stations.

The result obtained with the estimation of the seasonal component indicates a range of about $4\degree C$ between seasons, a result consistent with the findings previously reported in the literature (e.g.,   \cite{Stape2013b}, \cite{Iracema2009}). It is also possible to note that the estimated seasonal pattern is quite stable, consistent with the high estimated precision value. The estimated AR(2) component is consistent with a cyclic pattern. Partial autocorrelation coefficients were estimated with posterior means of 0.2891 and -0.046, which correspond to AR(1) and AR(2) coefficients of 0.3023 and -0.046, which have complex roots, generating a cycle period of 7.97 quarters. The cyclic component has a range of variation of about $2\degree C$, an important source of variability to explain the temperature patterns in this region.

%> roots= inla.ar.pacf2phi(c(resulttu$summary.hyper[5,1],resulttu$summary.hyper[6,1]))
%> (roots[1]^2+4*roots[2])
%[1] -0.02161895
%> (roots[1]^2+4*roots[2])<0
%[1] TRUE
%> f0=(1/(2*pi))*acos(roots[1]/(2*sqrt(-roots[2])))
%> 1/f0
%[1] 16.10966
%> inla.ar.pacf2phi(c(resulttu$summary.hyper[5,1],resulttu$summary.hyper[6,1]))
%[1]  0.35767235 -0.03738711

\begin{figure}
\caption{ \textbf{Spatial random effects - Average  temperature}\label{remean}}
\begin{center}
\subfloat[Estimated Random Effects]{\includegraphics[width=90mm,angle=270]{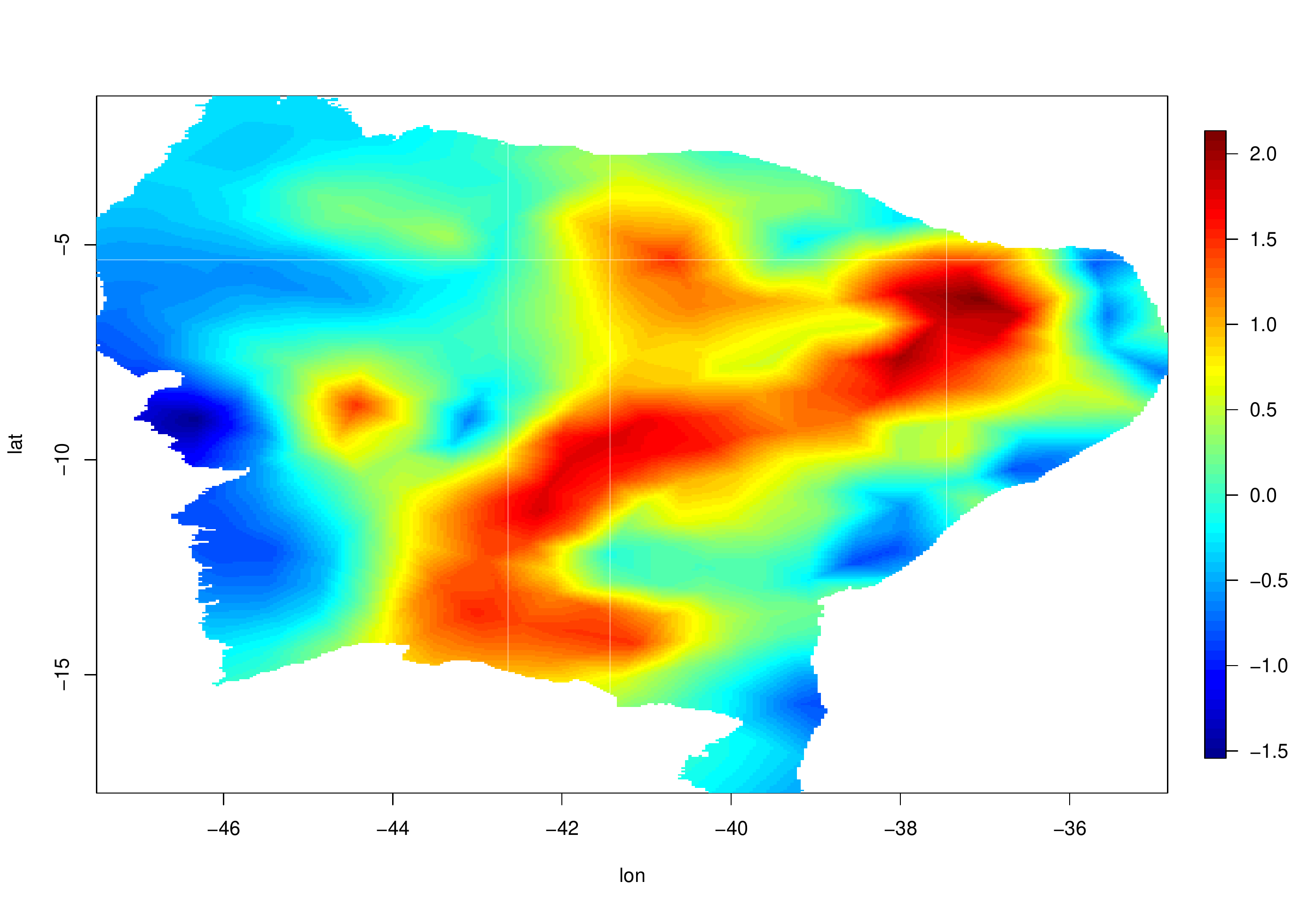}}\\
\subfloat[Estimated Random Effects and  K{\"o}ppen Climate Classification]{\includegraphics[width=90mm,angle=270]{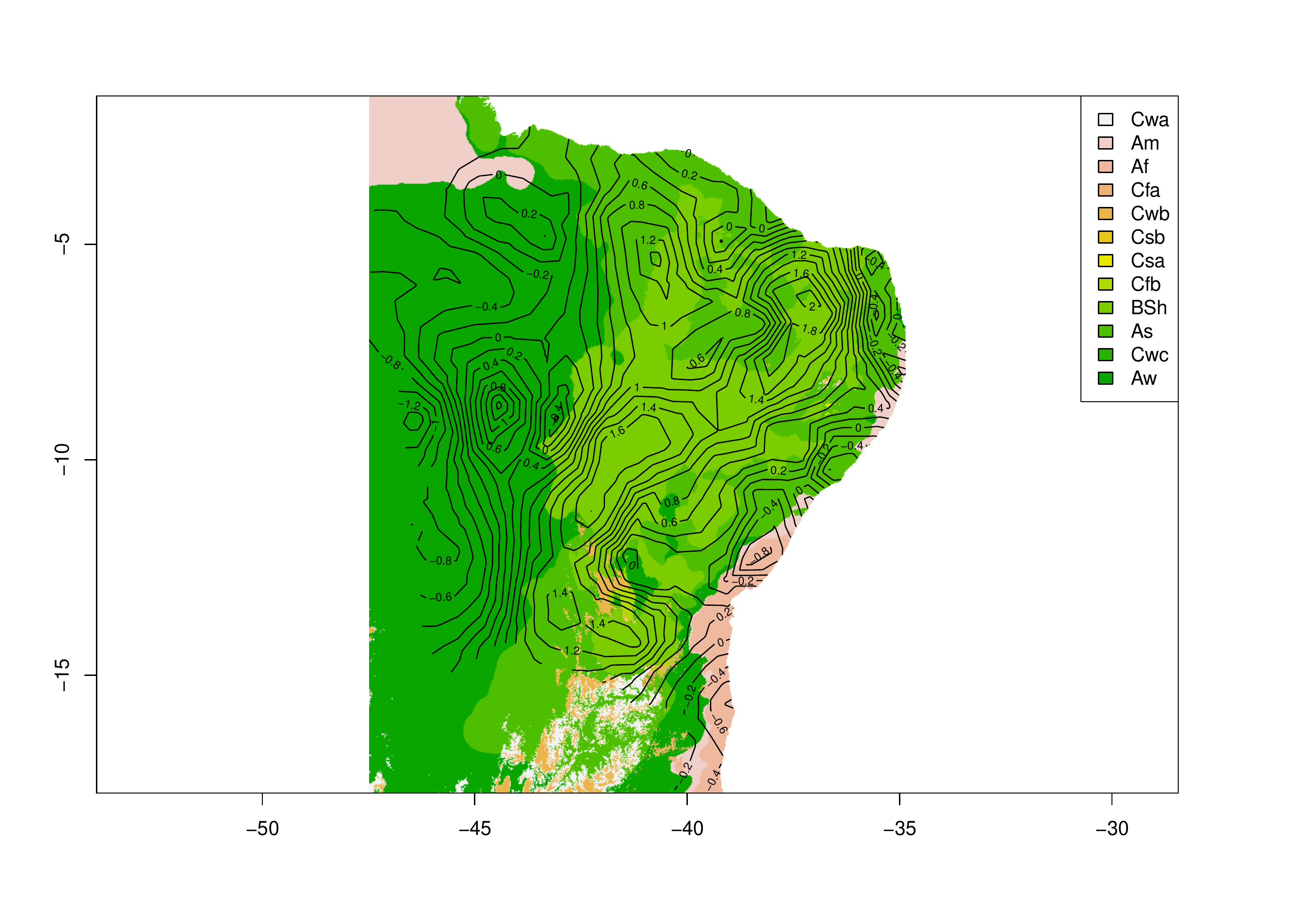}}\\
\vspace*{0.0cm}
\end{center}
\begin{footnotesize}
Note:  Posterior mean of estimated spatial random effects. 
\end{footnotesize}
\end{figure}

The importance of the spatial component in the explanation of the heterogeneity of temperatures can be seen in Figure \ref{remean}, which shows the spatial random effects estimated for the spatial continuum of the Northeast region. The first graph shows the spatial random effects, and the second figure shows these same effects placed as a contour plot over the K{\"o}ppen climate classification for this region. It is possible to observe that random effects capture with great precision the variability observed in this region, especially the higher average temperature observed in the semi-arid climate (Aw). It also captures the milder climates observed on the coast, in the forest zone and in the vicinity of Amazon forest. The amplitude of spatial effects is about $3.5\degree C$, a result consistent with the climatic variability observed in the Northeast region of Brazil.

\begin{figure}
\caption{ \textbf{Posterior Standard Deviation - Spatial random effects -   Average Temperature}\label{remeansd}}
\begin{center}
\includegraphics[width=90mm,angle=270]{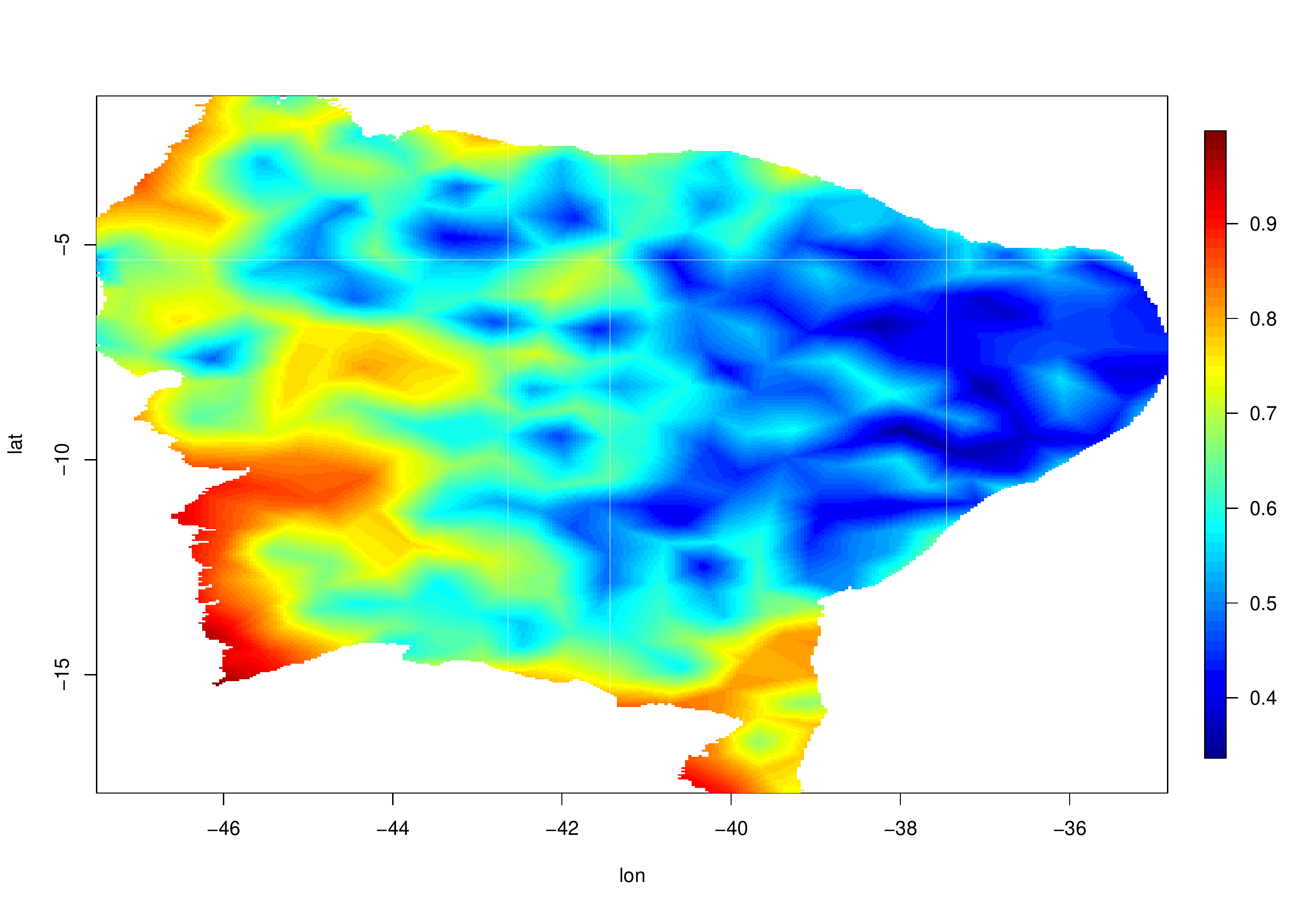}\\
\vspace*{0.0cm}
\end{center}
\end{figure}

It is important to note that the method used in this work allows us to construct uncertainty measures also for the estimated spatial random effects. Figure \ref{remeansd}  depicts the posterior standard deviations of the spatial random effects component reported in Figure  \ref{remean}, and equivalently we can construct credibility intervals for this measure. The values in this figure are associated with the density of weather monitoring stations in space, with a higher density of stations leading to more precise estimates.

\begin{figure}
\caption{ \textbf{Theoretical Matérn correlations - Average temperature }\label{teomatern}}
\begin{center}
\includegraphics[width=90mm,angle=270]{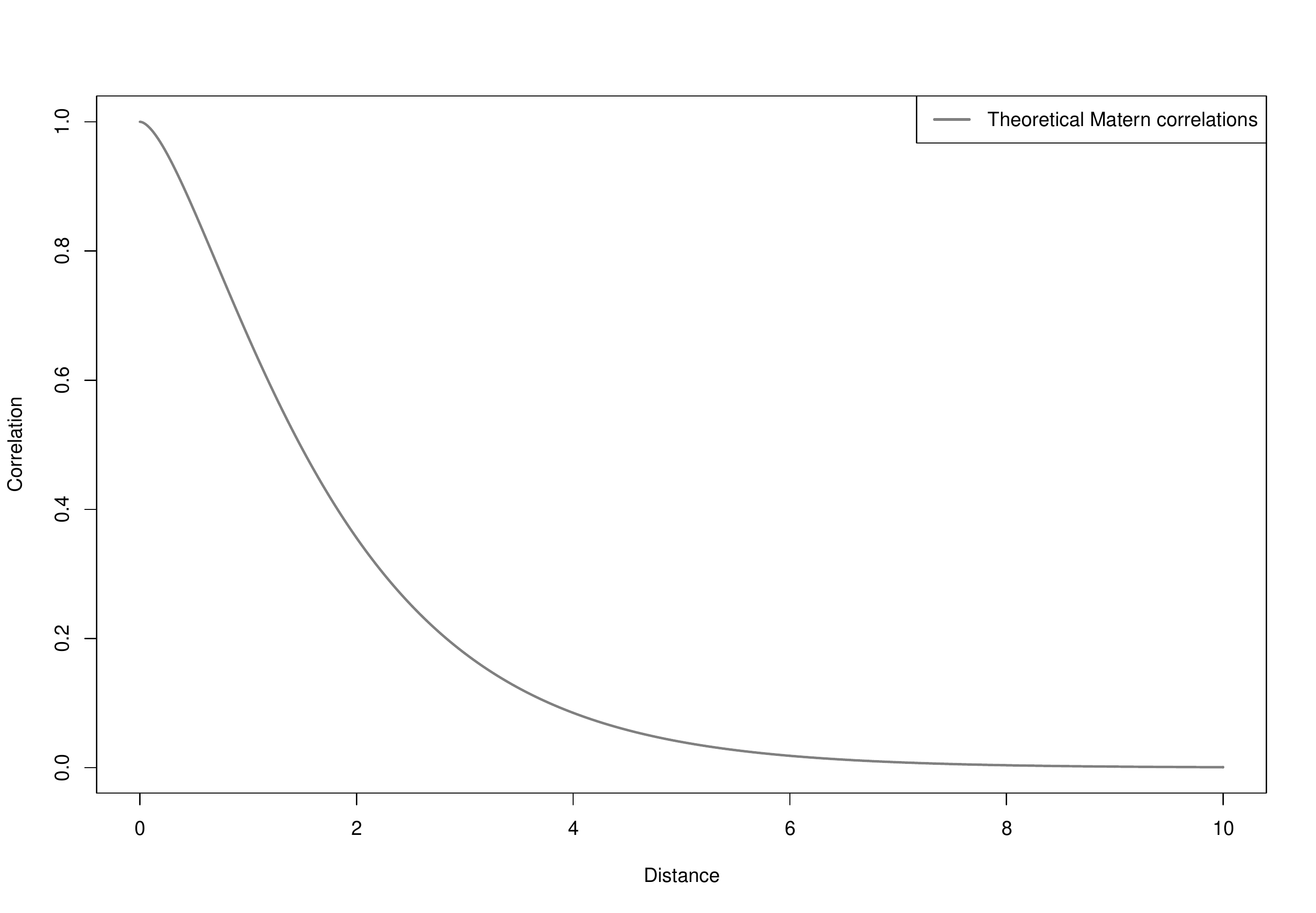}\\
\vspace*{0.0cm}
\end{center}
\begin{footnotesize}
Note:  Results based on the posterior mean estimated of log $\tau$ and log $\kappa$. 
\end{footnotesize}
\end{figure}

The spatial correlation pattern adjusted by the model can be observed in Figure \ref{teomatern}, which shows the spatial correlation function as a function of distance, and is a simpler way of interpreting the parameters  log $\tau$ and log $\kappa$ estimated by the model. It can be seen that the spatial correlation pattern is consistent with the spatial nature of climate effects, with neighboring regions having similar (correlated) climate patterns.

\begin{figure}
\caption{ \textbf{Fitted average temperature - 2014/4}\label{prevtemp20144}}
\begin{center}
\includegraphics[width=90mm,angle=270]{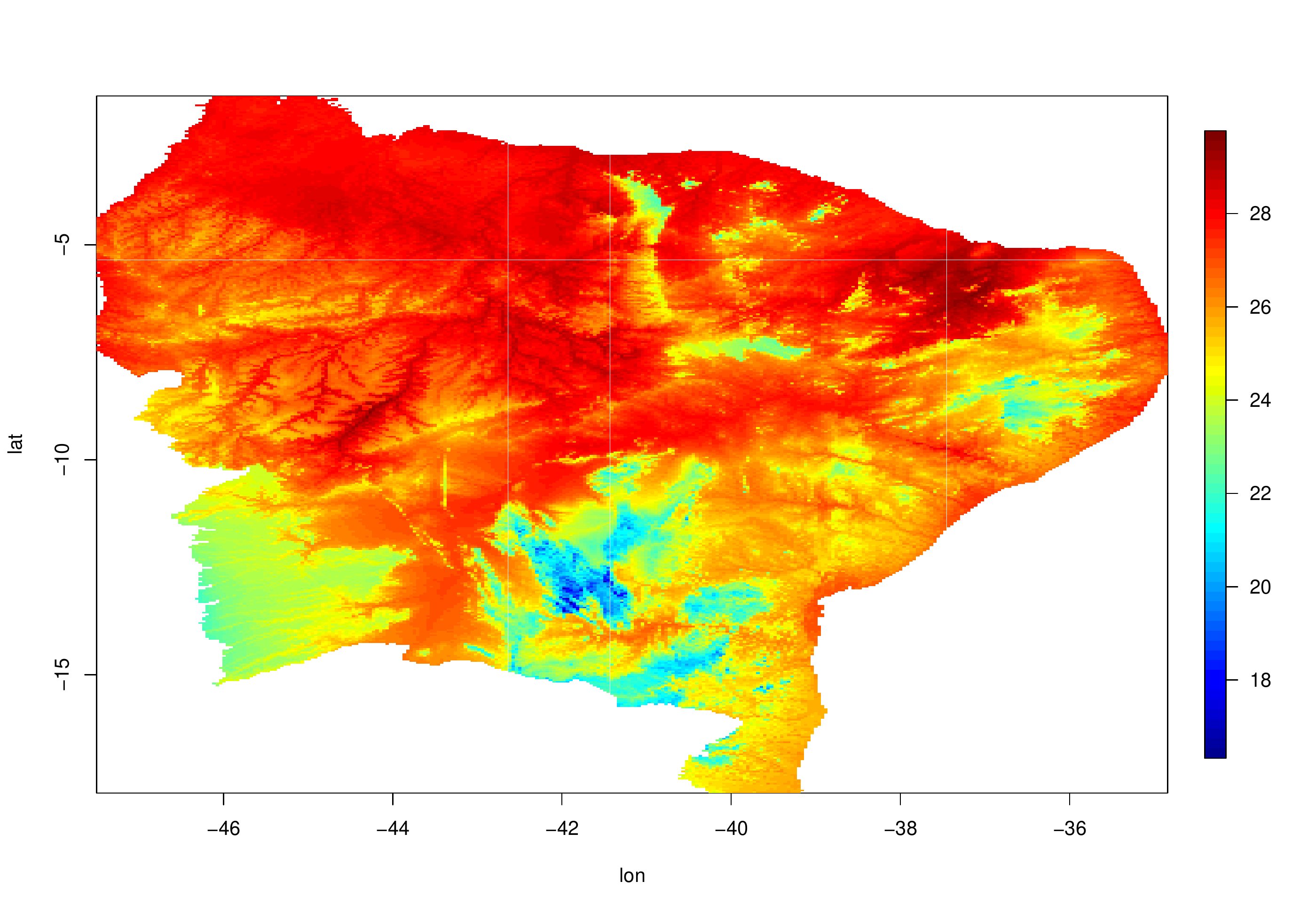}\\
\vspace*{0.0cm}
\end{center}
\begin{footnotesize}
Note:  Posterior mean of model fitted average temperature.  
\end{footnotesize}
\end{figure}

To show the model's ability to fit the temperature for the entire space analyzed, Figure \ref{prevtemp20144} shows the average temperature fitted  by the model for the last quarter of 2014 for the Northeast region of Brazil. This adjustment is constructed by adding the estimated posterior mean of the common components of trend, cycle and seasonality, the effects predicted by the explanatory variables for each point in the spatial continuum and the estimated spatial random effects. The model achieves a very satisfactory adjustment of the temperature variability in the Northeast region, being able to explain both the temperature patterns in the hottest semi-arid regions and near the Amazon forest, as well as the milder temperature regions such as the Chapada Diamantina in the center of Bahia and the Borborema Plateau, where the effects of altitude notably reduce the temperature.

In order to verify the importance of including the spatial components in this problem, we performed the estimation of the model without the inclusion of the spatial random effects structure. This estimation resulted in a marginal log-likelihood with a value of -19953.53, and a deviance information criterion of 39550.79 indicating a lower fit for the model with no spatial effects\footnote{The remaining results of this estimation are not essential and are therefore not reported, but are available on request.}. An important point is that the inclusion of the random spatial effect is fundamental for the correct identification of the components of trend, seasonality and cycle, and the estimation of the correct uncertainty associated with these components. 

\begin{figure}
\caption{ \textbf{Trend comparison}\label{trendcomparation}}
\begin{center}
\subfloat[Trend Comparison - Non Spatial vs Spatial]{\includegraphics[width=73mm]{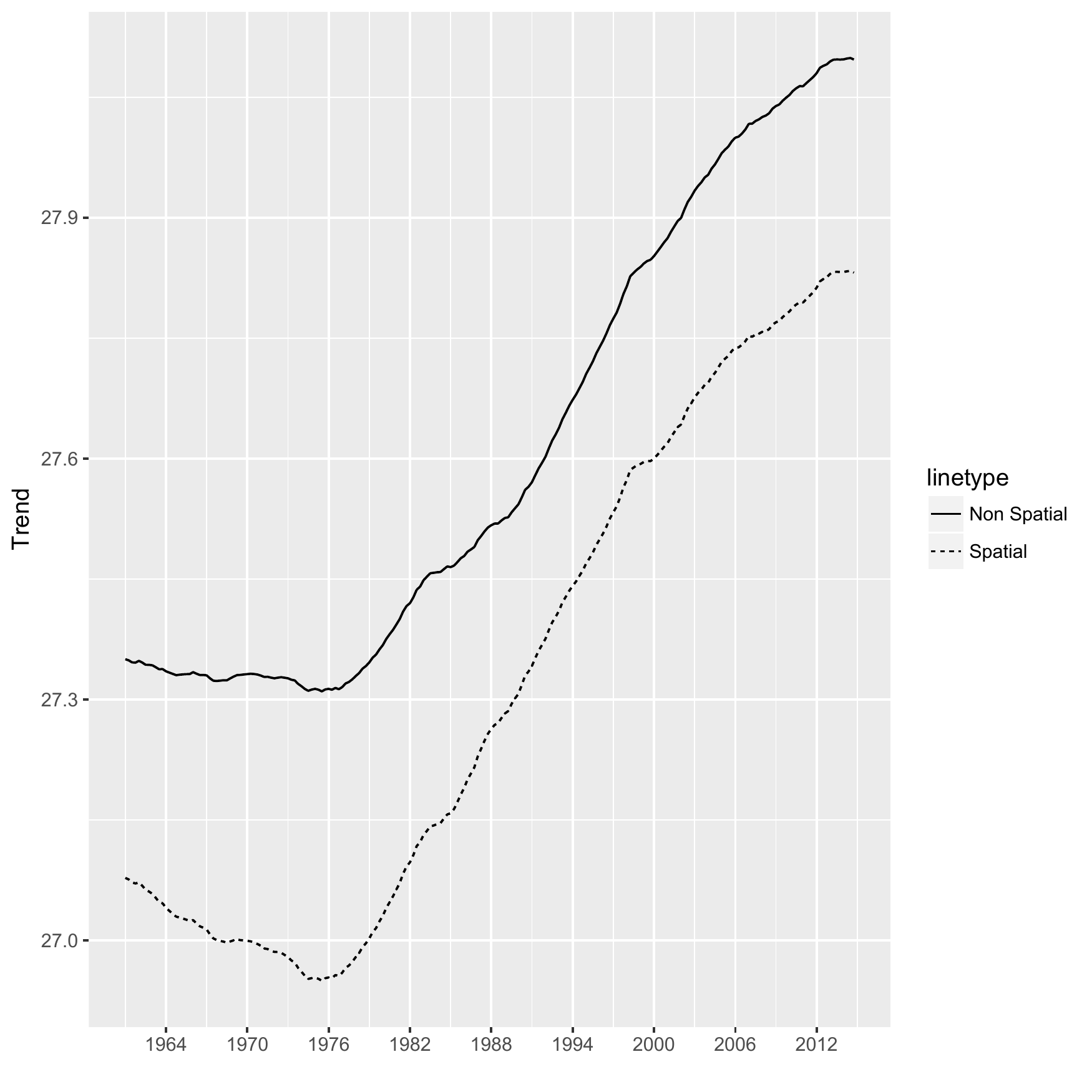}}\subfloat[Confidence IC - Trend for Non Spatial Model]{\includegraphics[width=73mm]{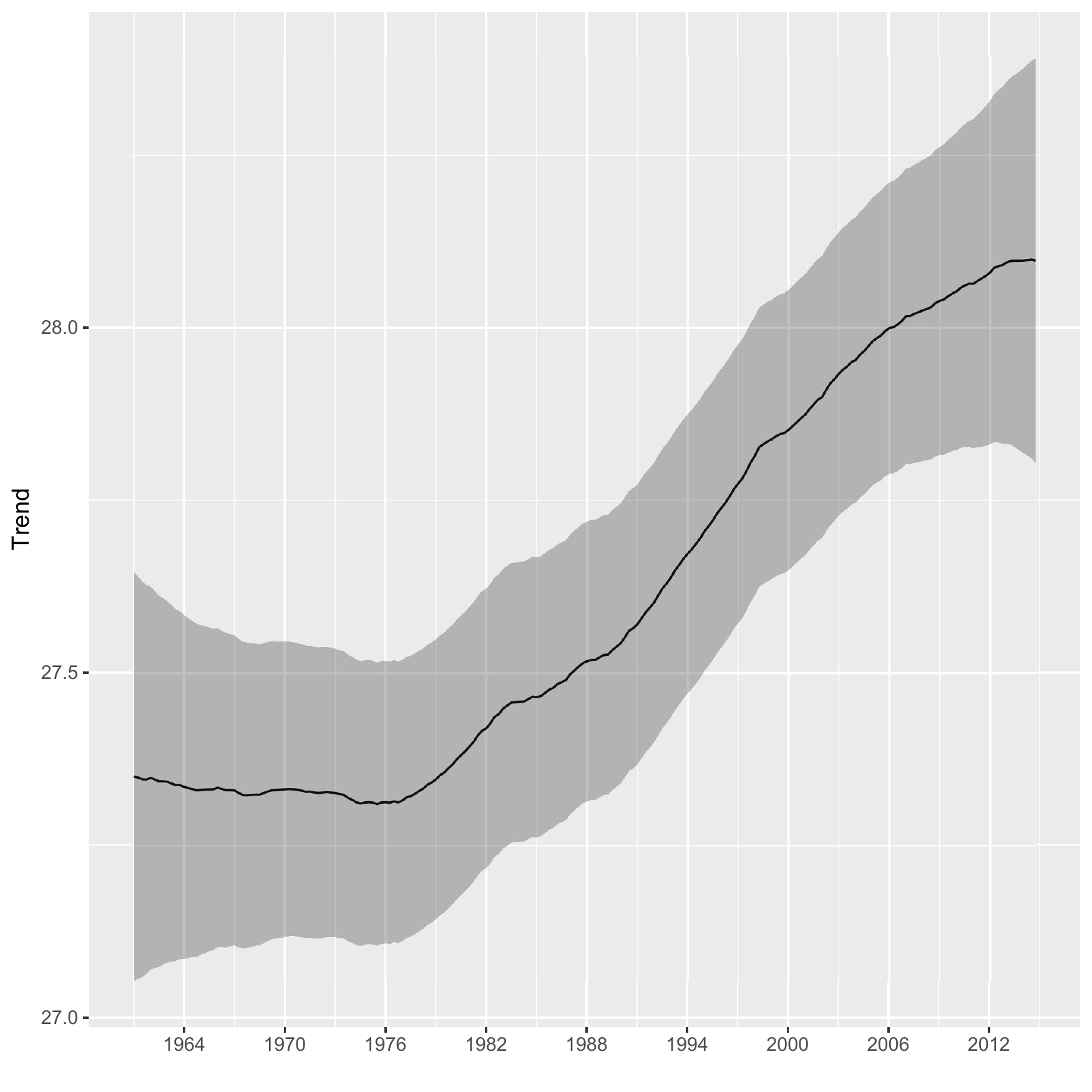}}\\
\vspace*{0.0cm}
\end{center}
\end{figure}

 In Figure \ref{trendcomparation}  we show the posterior mean values of the trend component for the models with and without the inclusion of random spatial effects, and the 95\% credibility interval for the trend component estimated by the model without the inclusion of the spatial random effects. It can be observed in sub-figure a) that the trend estimated by the model without spatial components is estimated with values greater than the trend estimated by the model with the inclusion of spatial component. Another problem is indicated by sub-figure b), which shows that the width of the credibility interval is smaller than that estimated with the inclusion of the spatial random effects (sub-figure a) of Figure  \ref{bsmmean}). This shows that the model without spatial effects does not capture the cross-sectional spatial heterogeneity existing in the temperatures, and thus does not allow correctly recovering the uncertainty in the estimation of this component. These results indicate that the inclusion of the spatial components is important for the correct estimation of the components and their variability in this context.

\subsection{Alternative specification with constant mean}

\begin{figure}
\caption{ \textbf{AR(2) Component  - Constant mean specification - Average Temperature}\label{ar2cm}}
\begin{center}
\includegraphics[width=90mm,angle=0]{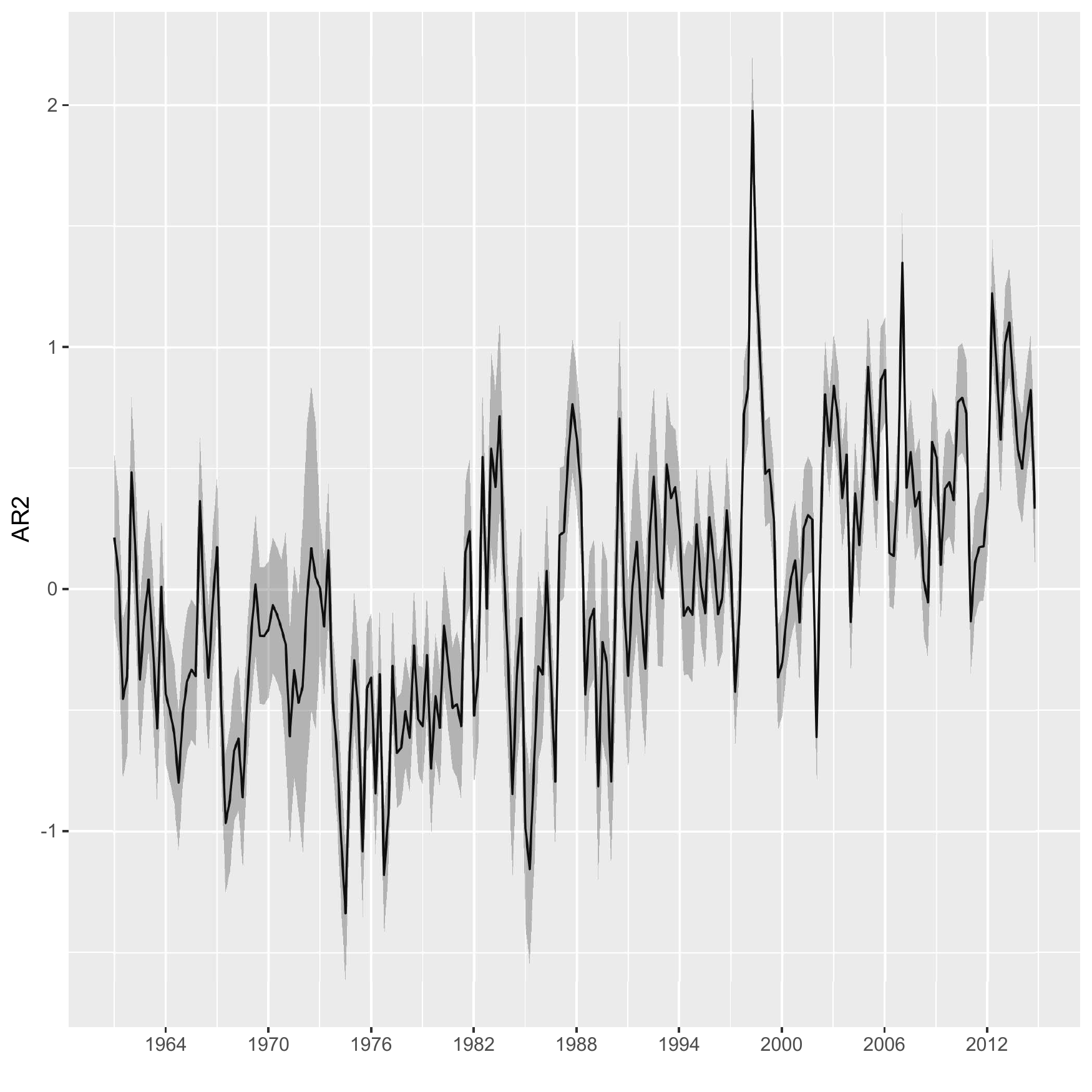}\\
\vspace*{0.0cm}
\end{center}
\end{figure}

The use of the specification based on a random walk model for the trend follows the specification adopted in other works, e.g., \cite{Gordon1991}. But there is a deeper discussion about the use of this specification in climate data modeling, especially whether these series are in fact stationary or non-stationary, as analyzed in \cite{Zheng1999}, \cite{Kaufmann2006} and  \cite{Proietti2017}. 
According to our knowledge, there is no formal test, such as a spatial unit root test or a test for trend specification (e.g., \cite{Nyblon1986}), adequate for the context of spatio-temporal model with continuous spatial random effects analyzed in our work. To analyze this issue, we estimated an alternative specification, reported in Table  \ref{tab:constantmean}, assuming a constant mean plus seasonality and AR(2) processes, which imposes a stationarity structure for the temperature series. This alternative specification obtains a DIC of 35677.76 and marginal likelihood of -18179.13, being for these criteria a specification inferior to the model report in Table \ref{tab:mean}. The estimated AR(2) component has a rather distinct dynamic compared to the previous result, with PACF1 and PACF2 with values of 0.6738 and 0.1004, corresponding to AR coefficients of 0.6057  and 0.1004, which does not have complex roots and thus without the presence of cycles. Figure \ref{ar2cm} shows the evolution of this component. We can see that in this model the temperature rise at the end of the sample is captured by the positive values for this component.

By the adjustment results, the model with the trend specification seems to result in a more adequate specification for this data set. The trend specification based on the random walk model is also consistent with the trend definition used in climate analyzes, as discussed by  \cite{Gordon1991} and \cite{Ji2014}, and thus we have adopted this specification in this paper, but we recognize that this question is a complex and central issue in the analysis of climate change.

\begin{table}
\protect\caption{\label{tab:constantmean}\textbf{Estimated parameters -  Constant mean specification - Average temperature} }
\begin{small}
\begin{tabular}{ccccccc}
\hline 
 & {\small{}mean} & {\small{}sd} & {\small{}.025q} & {\small{}.5q} & {\small{}.975q} & {\small{}mode}\tabularnewline
\hline 
{\small{}Mean Temp.} &  27.4197 &0.3521&    26.7284 & 27.4197&    28.1104 &27.4197 \tabularnewline
{\small{}Altitude} &   -0.0074 &0.0002  &  -0.0079&  -0.0074 &   -0.0069 &-0.0074 \tabularnewline
{\small{}Latitude } &  0.1301 &0.0364   &  0.0587&   0.1301 &    0.2015 & 0.1301  \tabularnewline
{\small{}Distance to Sea } &  0.0047 &0.0006&     0.0034&   0.0047   &  0.0059 & 0.0047  \tabularnewline
Precision Gaussian &  0.9652 &1.240e-02   &  0.9410   &  0.9651   &  0.9898 &   0.9650\tabularnewline
Precision Seasonal  &   24376.7858& 2.165e+04&  2499.9460& 18413.2611 &81447.7425 &7224.6054\tabularnewline
Precision Cycle  & 3.1781& 5.713e-01 &    2.2301 &    3.1170  &   4.4671&    2.9915\tabularnewline
PACF1  &  0.6738 &6.240e-02 &    0.5325&     0.6810 &    0.7760  &  0.6971\tabularnewline
PACF2 &    0.1004 &7.800e-02 &   -0.0421&     0.0962  &   0.2620 &   0.0823\tabularnewline
log $\tau$ &     -1.3779& 1.195e-01  &  -1.6344   & -1.3761  &  -1.1313   &-1.3705\tabularnewline
log $\kappa$ &   0.3834 &1.268e-01 &    0.1280&     0.3801&     0.6587&    0.3755\tabularnewline
Marginal Lik. &-18179.13   & obs  & 12323 &  &  & \tabularnewline
DIC &  35677.76& &   &  &  & \tabularnewline
\hline 
\end{tabular}
\end{small}
\begin{tiny}\\
\end{tiny}
\end{table}

\subsection{Results - Maximum Temperature}

We also present the results obtained from estimating the model for the maximum temperature series, presented in Table \ref{tab:maximum}. In general, the results are similar to those obtained for the average temperature, with higher values. The estimated trend, seasonality and cycle components (Figure \ref{tab:maximum}) are analogous to those obtained for the average temperature, and especially the trend component shows that the maximum temperature in this region also shows a relevant elevation from 1976, indicating that the thermal amplitude also increased in the last 30 years.

\begin{table}
\protect\caption{\label{tab:maximum}\textbf{Estimated parameters - Maximum temperature} }
\begin{small}
\begin{tabular}{ccccccc}
\hline 
 & {\small{}mean} & {\small{}sd} & {\small{}.025q} & {\small{}.5q} & {\small{}.975q} & {\small{}mode}\tabularnewline
\hline 
{\small{}Altitude} &   -0.0078& 0.0004 &   -0.0086 & -0.0078  &  -0.0071& -0.0078 \tabularnewline
{\small{}Latitude } &  0.1505& 0.0709  &   0.0112 &  0.1505  &   0.2896 & 0.1505   \tabularnewline
{\small{}Distance to Sea } &  0.0094& 0.0012 &    0.0070&   0.0094&     0.0118 & 0.0094   \tabularnewline
Precision Gaussian &  0.2014& 2.500e-03 &    0.1965 &    0.2014&  2.063e-01 &   0.2014\tabularnewline
Precision RW &714.3660& 6.787e+02 &  130.0317 &  514.4938&  2.491e+03 & 297.3961\tabularnewline
Precision Seasonal  &   30898.3910& 3.964e+04&  3044.1120& 19028.2506 & 1.314e+05 &7824.0889\tabularnewline
Precision Cycle  & 2.9121& 3.570e-01   &  2.2438  &   2.9039 & 3.643e+00 &   2.8990\tabularnewline
PACF1  &  0.3778& 7.970e-02 &    0.2054  &   0.3844  &5.165e-01 &   0.4014\tabularnewline
PACF2 &    -0.1003& 8.530e-02 &   -0.2763  &  -0.0960 & 5.670e-02&   -0.0817\tabularnewline
log $\tau$ &    -1.6363& 1.546e-01  &  -1.9599  &  -1.6312& -1.334e+00&   -1.6203\tabularnewline
log $\kappa$ &   0.1170& 1.690e-01  &  -0.1982  &   0.1117 & 4.761e-01  &  0.0995\tabularnewline
Marginal Lik. &-27719.46   & obs  & 12323 &  &  & \tabularnewline
DIC & 54957.47& &   &  &  & \tabularnewline
\hline 
\end{tabular}
\end{small}
\begin{tiny}\\
\end{tiny}
\end{table}

\begin{figure}
\caption{ \textbf{Trend, seasonal and cycles decomposition - Maximum temperature }\label{bsmmaximum}}
\begin{center}
\subfloat[Trend]{\includegraphics[width=80mm]{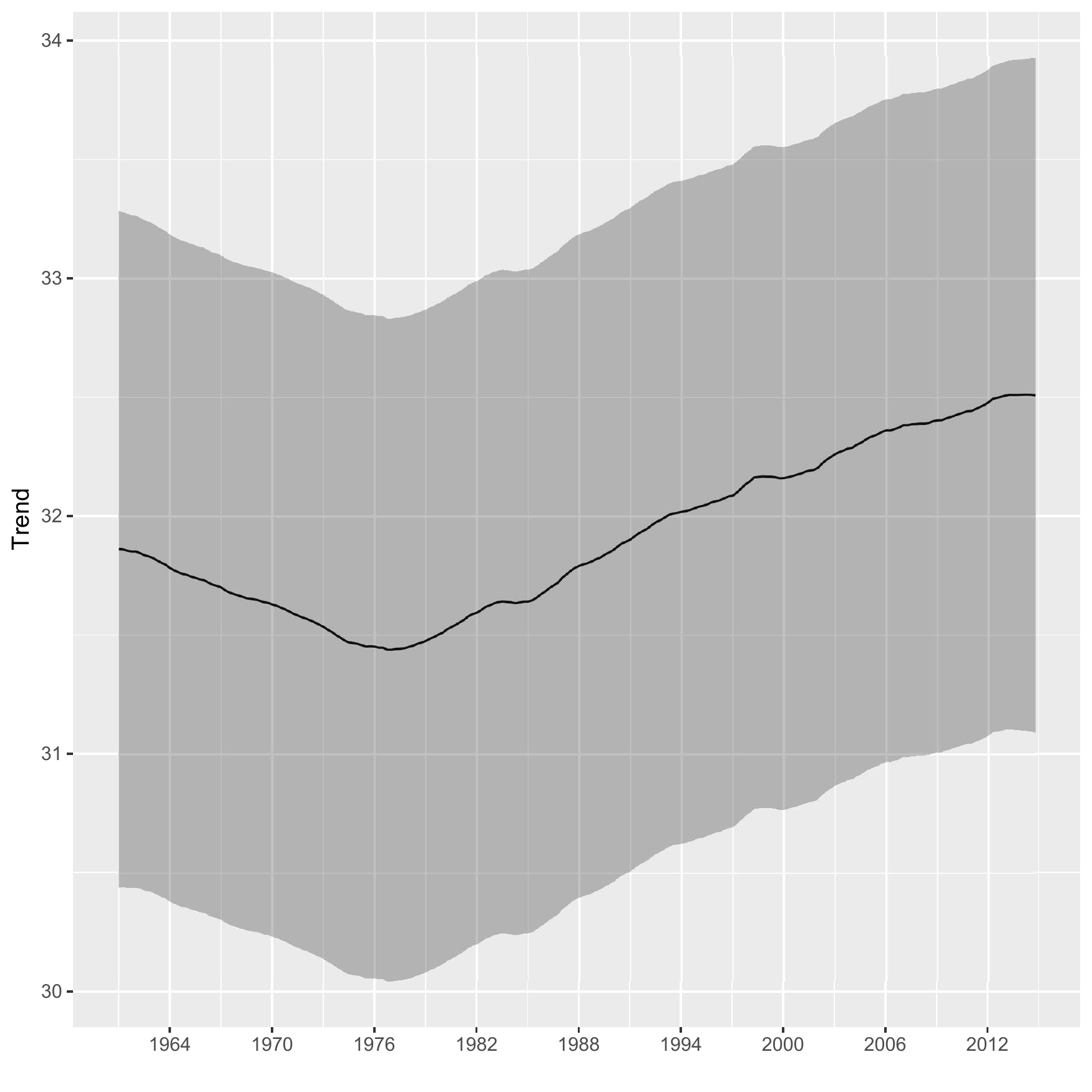}}\subfloat[Seasonal]{\includegraphics[width=80mm]{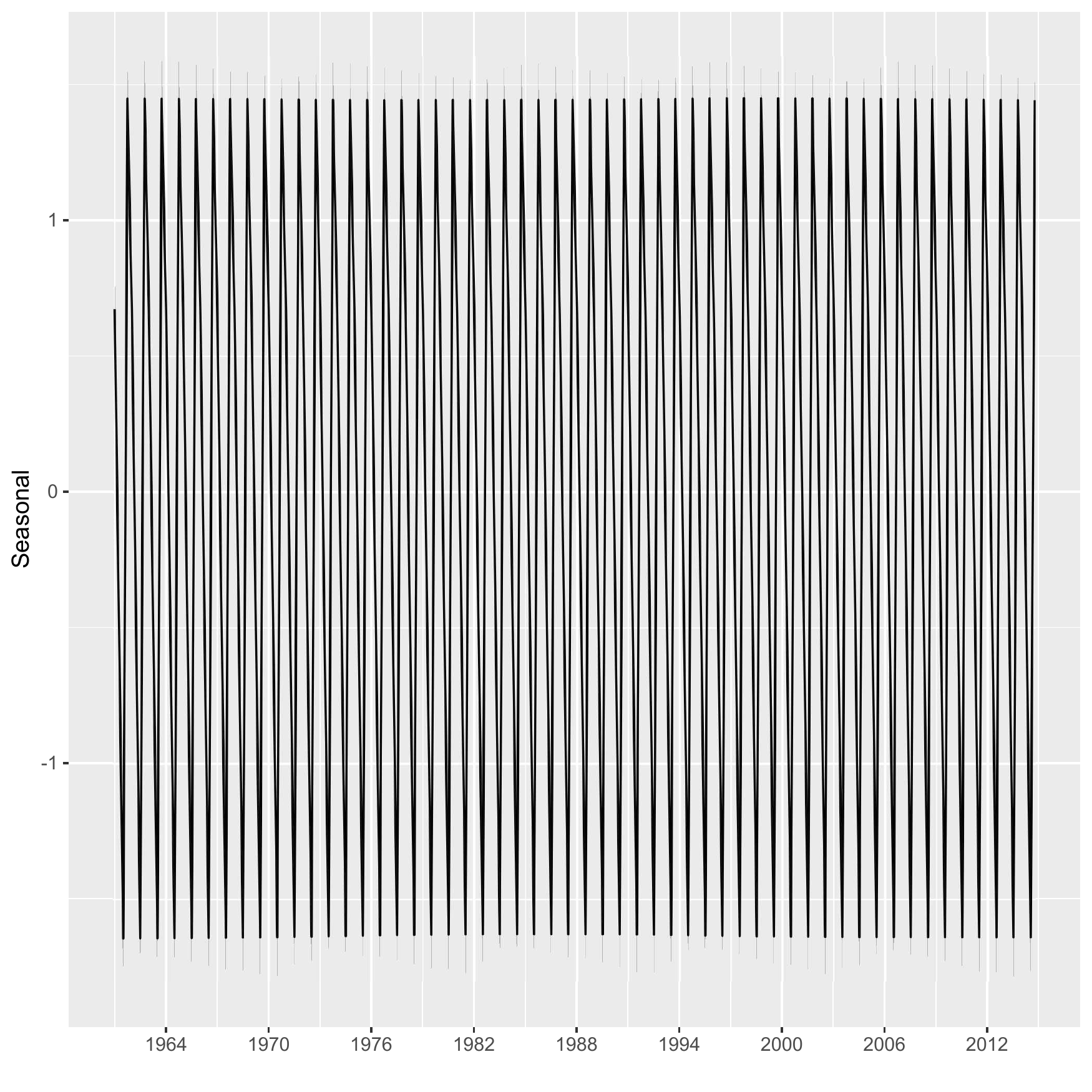}}\\
\subfloat[Cycle]{\includegraphics[width=80mm]{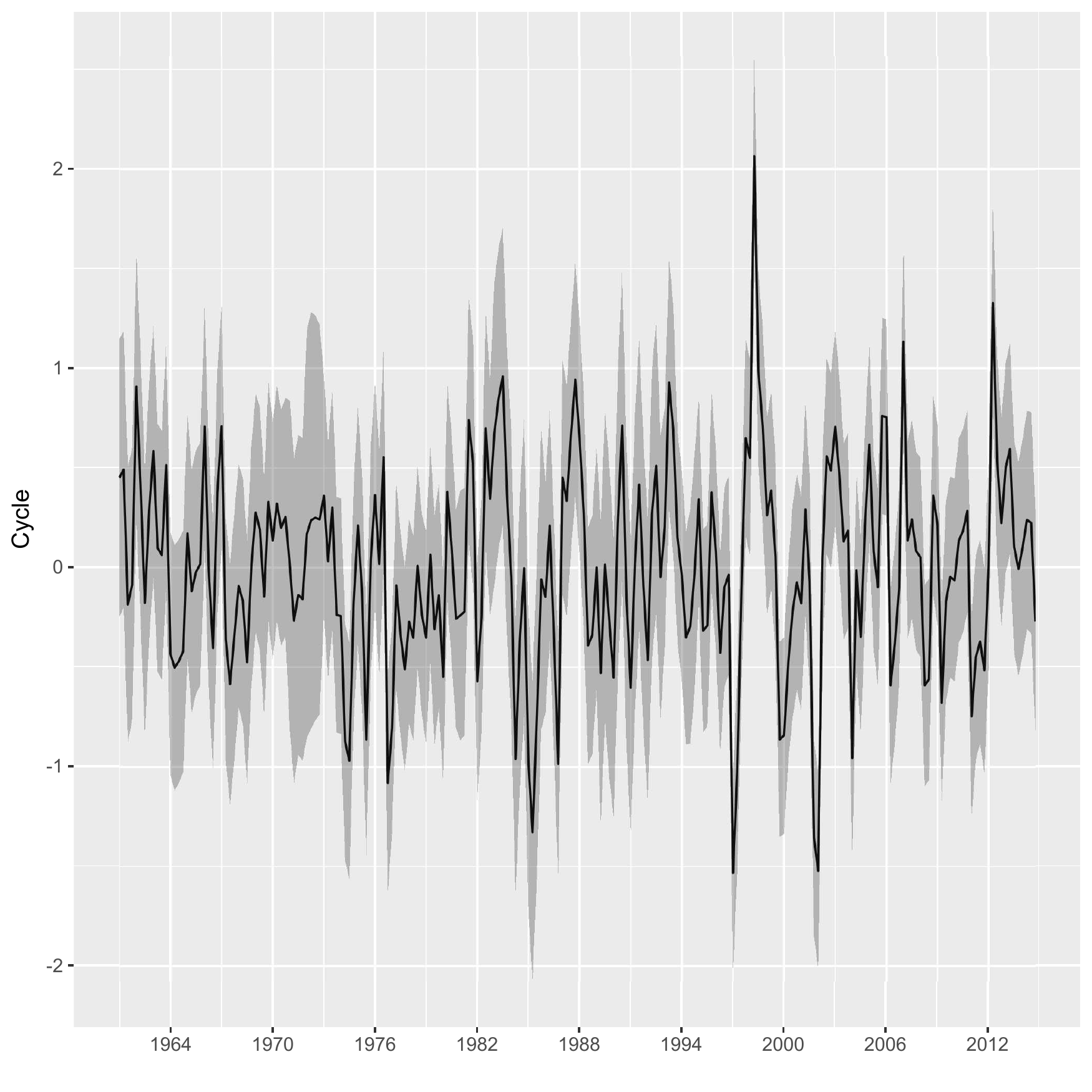}}
\vspace*{0.0cm}
\end{center}
\end{figure}

\begin{figure}
\caption{ \textbf{Fitted Maximum Temperature}\label{prevtempmaxima20144}}
\begin{center}
\includegraphics[width=90mm,angle=270]{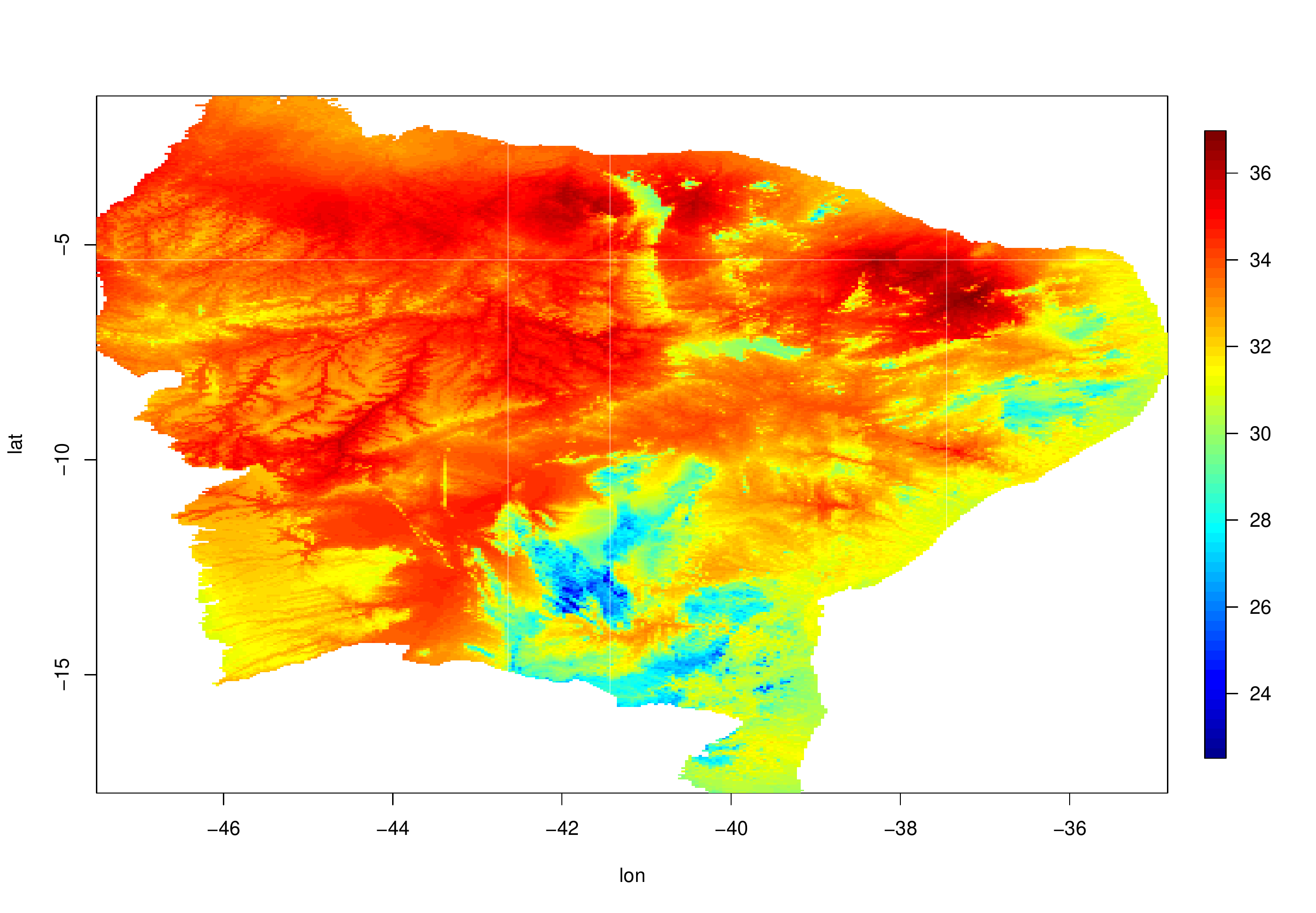}\\
\vspace*{0.0cm}
\end{center}
\begin{footnotesize}
Note:  Posterior mean of model fitted  maximum temperature.  
\end{footnotesize}
\end{figure}

 Figure \ref{bsmmaximum} shows the spatial random effects estimated for the maximum temperature. It is possible to note that this process has  larger amplitude than that observed for the average temperature, being consistent with the patterns of temperature variability observed in Northeast climates, as discussed in \cite{Stape2013}. The adjustment of the model to the maximum temperature in the Northeast region in 2014/4 is shown in Figure \ref{prevtempmaxima20144}, indicating the high maximum temperatures of this region during this period.

\begin{figure}
\caption{ \textbf{Spatial random effects - Maximum temperature}\label{remaxima}}
\begin{center}
\includegraphics[width=90mm,angle=270]{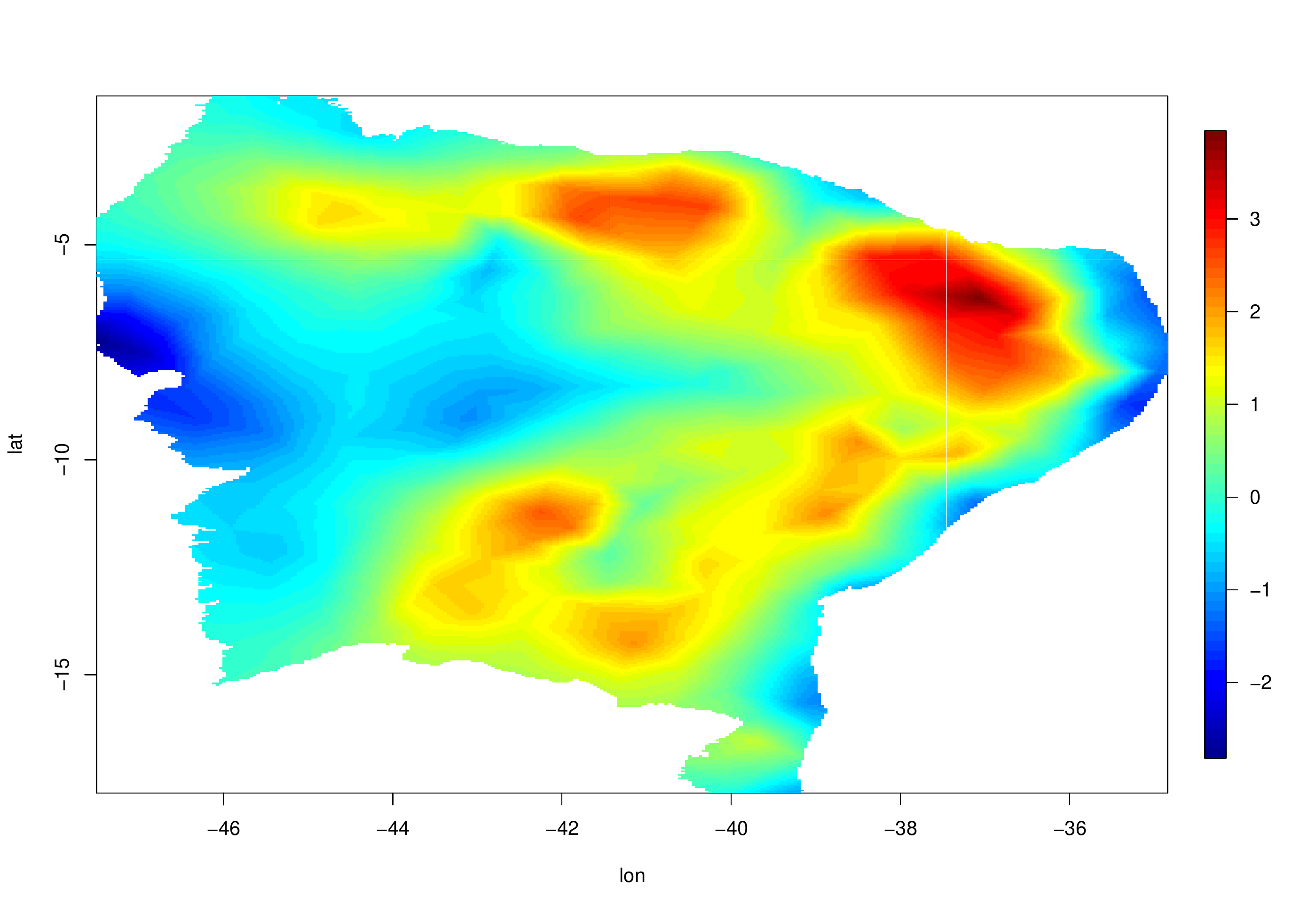}\\
\vspace*{0.0cm}
\end{center}
\begin{footnotesize}
Note:  Posterior mean of estimated spatial random effects. 
\end{footnotesize}
\end{figure} 

\subsection{Rainfall}

After modeling the temperature series, our objective is to show how this model can be applied to explain the accumulated rainfall. For this we model the quarterly series of rainfall, defined as the accumulated daily rainfall throughout the quarter. The Brazilian Northeast is especially sensitive to the problem of droughts, and the climate issue is one of the main barriers to the economic development of the region, which is the worst in terms of poverty and human development in Brazil. The issue of drought is also one of the fundamental aspects of Brazilian culture, being the theme of fundamental works in Brazilian literature such as Vidas Secas (Barren Lives) by Graciliano Ramos (\cite {ramos1938}) and Raquel Queiroz's The Fifteen (\cite{Queiroz1927}), literary portraits of the desolation of drought and hunger in Northeast Brazil.

\begin{table}
\protect\caption{\label{tab:rainfall}\textbf{Estimated parameters - Rainfall}}
\begin{small}
\begin{tabular}{ccccccc}
\hline 
 & {\small{}mean} & {\small{}sd} & {\small{}.025q} & {\small{}.5q} & {\small{}.975q} & {\small{}mode}\tabularnewline
\hline 
{\small{}Altitude} &   0.081 &0.0263 &    0.0293&    0.081 &    0.1327& 0.081    \tabularnewline
Precision Gaussian &  0.0000&  0.0000  &   0.0000 &   0.0000   &  0.0000  &  0.0000\tabularnewline
Precision RW &3281.5312 &74.4447 & 3135.3441 &3281.1874&  3429.4133 &3280.8979\tabularnewline
Precision Seasonal  &  57.4686&  1.3319 &   54.8634 &  57.4610 &   60.1045 &  57.4523\tabularnewline
Precision Cycle  &  0.0003 & 0.0000  &   0.0003 &   0.0003 &    0.0003 &   0.0003\tabularnewline
PACF1 &   0.3279 & 0.0096 &    0.3083 &   0.3281  &   0.3462   & 0.3288\tabularnewline
PACF2 &    -0.0716 & 0.0095  &  -0.0894  & -0.0719  &  -0.0518  & -0.0728\tabularnewline
log $\tau$ &    -5.4688 & 0.0211  &  -5.5177  & -5.4653   & -5.4368 &  -5.4563\tabularnewline
log $\kappa$ &   -1.6862 & 0.0097   & -1.7009  & -1.6879  &  -1.6636 &  -1.6925\tabularnewline
Marginal Lik. &-82544.41    & obs  & 12323 &  &  & \tabularnewline
DIC & 164351.23&  &   &  &  & \tabularnewline
\hline 
\end{tabular}
\end{small}
\begin{tiny}\\
Estimation results - Rainfall
\end{tiny}
\end{table}

Table \ref{tab:rainfall} presents the results of estimating the model. The chosen specification by the DIC uses only altitude as an explanatory variable. In this formulation, the use of the latitude and longitude variables led to convergence and identification problems, and the distance to the sea had no relevant explanatory power. The first notable aspect in the estimation for rainfall is the lower precision for the Gaussian component, which is consistent with the greater range of observed values for rainfall, and the smaller precision for the components of seasonality and cycle, consistent with the greater seasonal and inter-annual variability of rainfall in this region, according to \cite{Strang1972} and \cite{Molion2002}.

\begin{figure}
\caption{ \textbf{Trend, seasonal and cycles decomposition - Rainfall}\label{bsmpluv}}
\begin{center}
\subfloat[Trend]{\includegraphics[width=80mm]{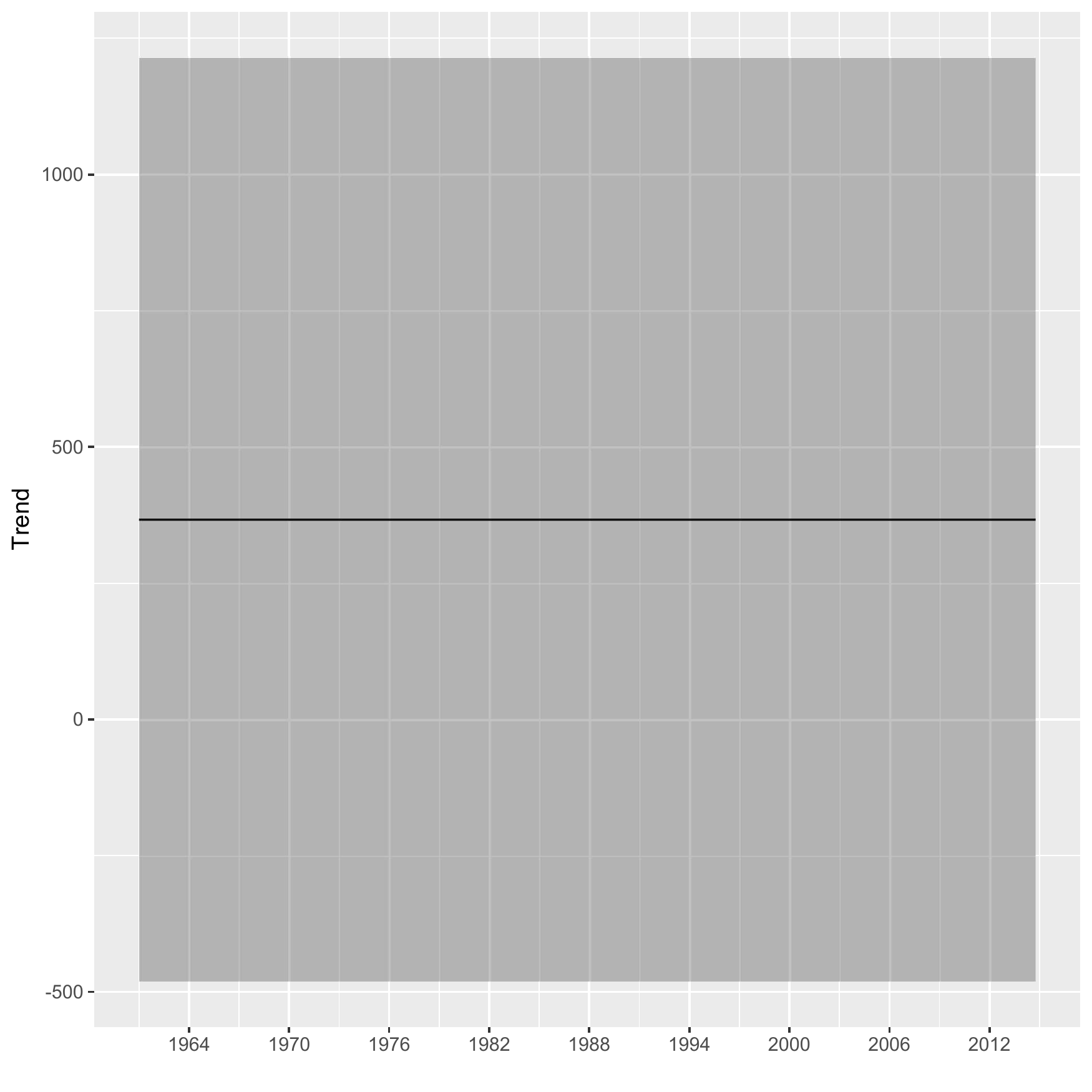}}\subfloat[Seasonal]{\includegraphics[width=80mm]{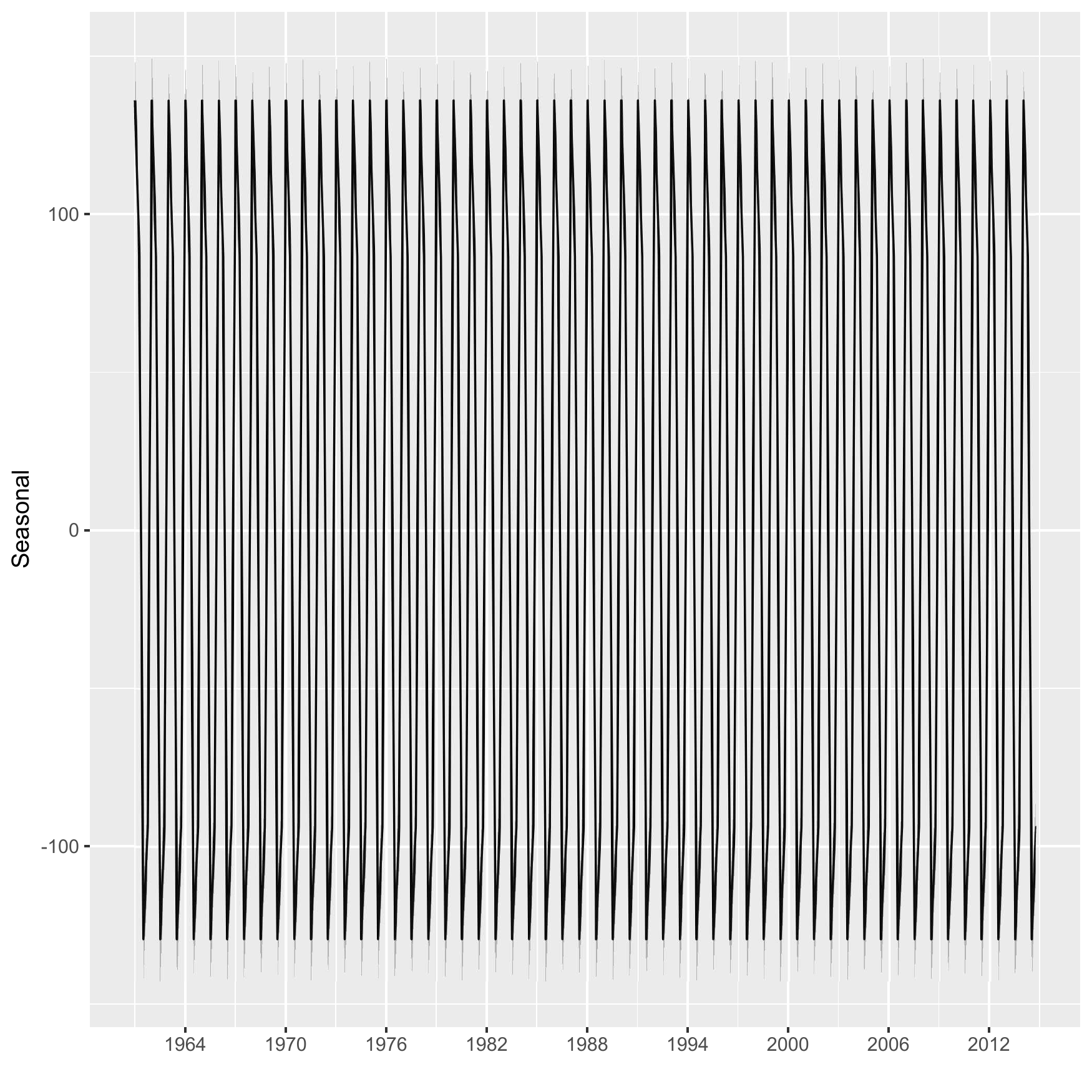}}\\
\subfloat[Cycle]{\includegraphics[width=80mm]{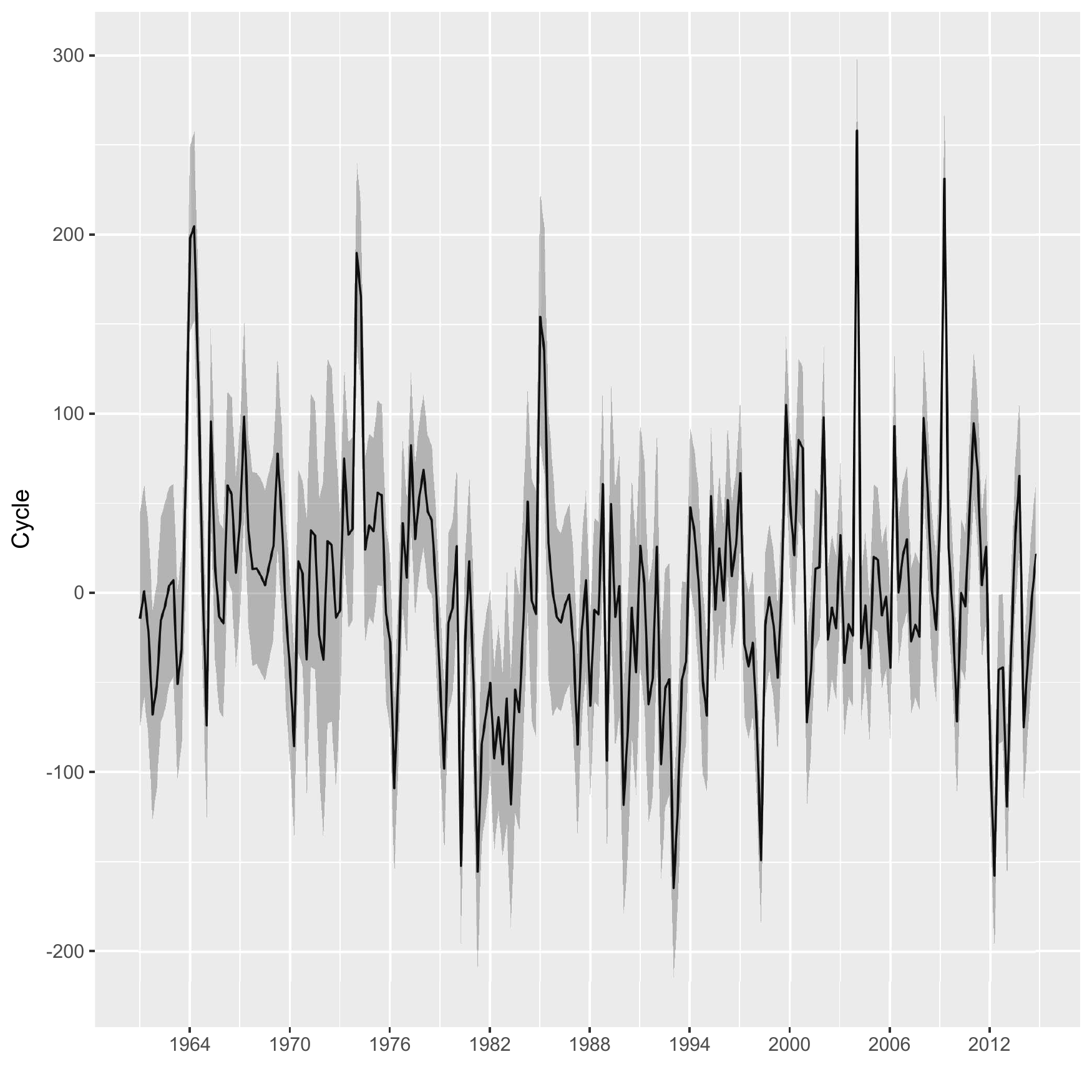}}
\vspace*{0.0cm}
\end{center}
\end{figure}

In Figure \ref{bsmpluv} we present the trend, seasonality and cycle components estimated for the rainfall series. The trend component is basically estimated as a constant, with a very wide credibility interval, which is consistent with the large inter-annual variability observed in this series. Contrary to the results obtained for the temperature series, it is not possible to obtain evidence of changes in the trend component for the rainfall series for the Northeast region of Brazil. The cyclical component was estimated with a frequency of 7.35 quarters, and presents very high amplitude, with values between 200 mm and -150 mm for some quarters, which is quite relevant, since the trend component is around 363 mm.

%> roots= inla.ar.pacf2phi(c(resulttu$summary.hyper[5,1],resulttu$summary.hyper[6,1]))
%> (roots[1]^2+4*roots[2])
%[1] -0.1628632
%> (roots[1]^2+4*roots[2])<0
%[1] TRUE
%> f0=(1/(2*pi))*acos(roots[1]/(2*sqrt(-roots[2])))
%> 1/f0
%[1] 7.353273

\begin{figure}
\caption{ \textbf{Spatial random effects -   Rainfall}\label{repluv}}
\begin{center}
\includegraphics[width=90mm,angle=270]{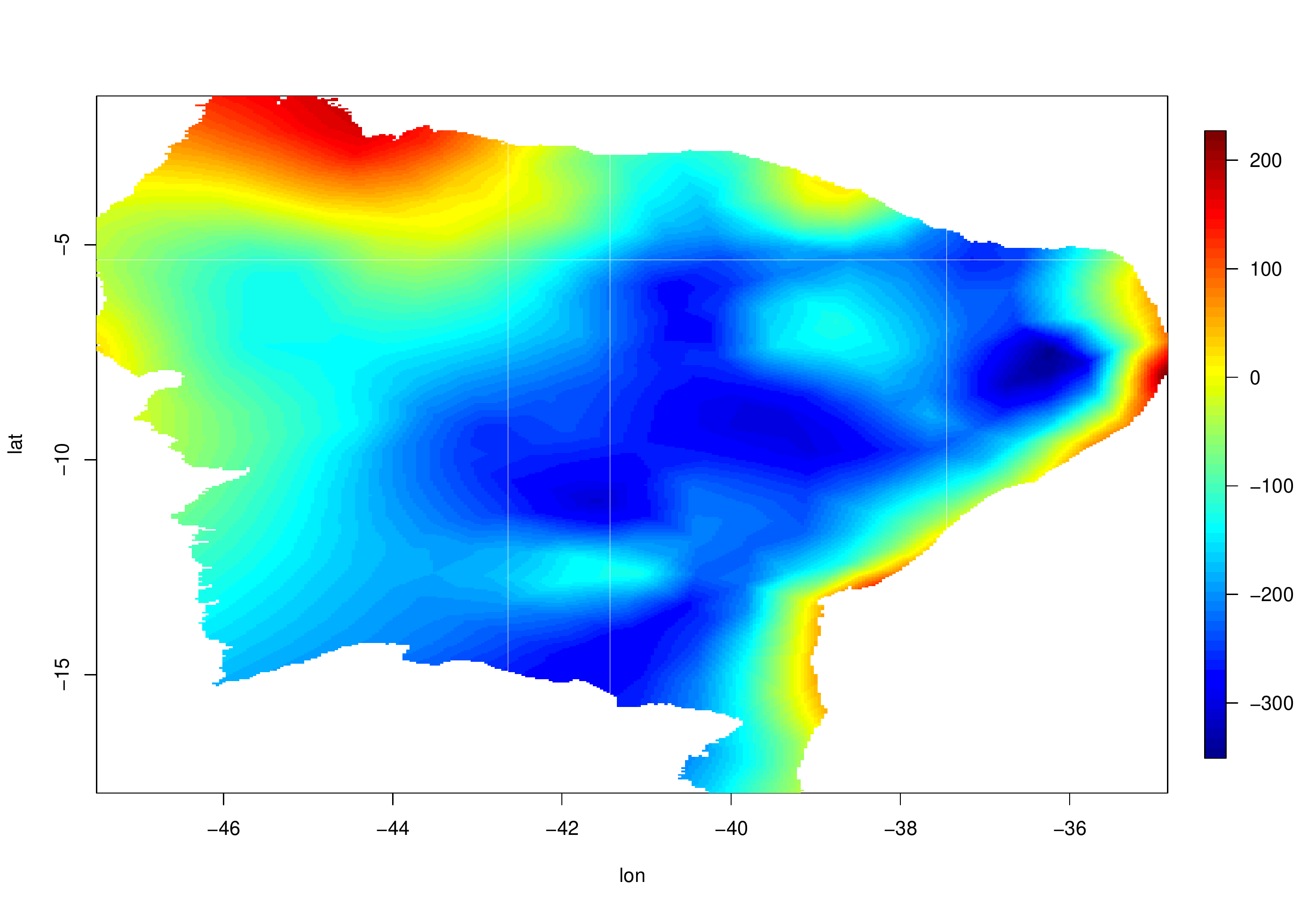}\\
\vspace*{0.0cm}
\end{center}
\begin{footnotesize}
Note:  Posterior mean of estimated spatial random effects. 
\end{footnotesize}
\end{figure}

A fundamental aspect in the climatic analysis of the Northeast region is the great spatial variability in the rainfall pattern. In this aspect, the model contributes significantly to the estimation of the continuous spatial random effect, which allows us to capture the fundamental spatial aspects in this question. Figure \ref{repluv} shows the estimate of the spatial random effects associated with rainfall. It is possible to observe that this component captures the high climatic heterogeneity observed in this region. This component shows the large rainfall deficit in the semi-arid regions, with negative patterns up to -300 mm in each quarter, consistent with periods with no rainfall, as well as the regions with the highest rainfall on the coast and in the northern region near the Amazon forest.

\begin{figure}
\caption{ \textbf{Fitted rainfall - 2001/3 }\label{prevpluv}}
\begin{center}
\includegraphics[width=90mm,angle=270]{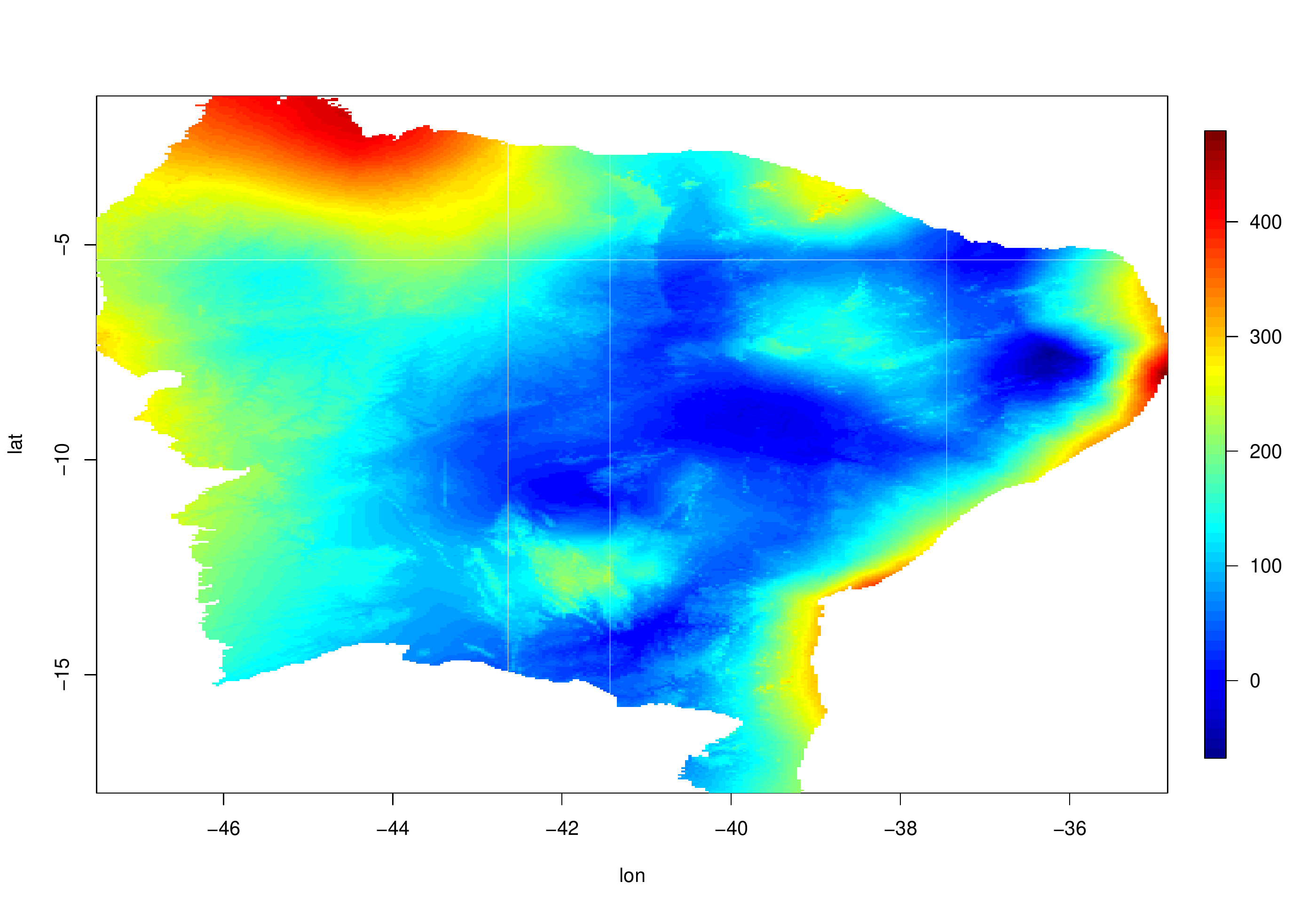}\\
\vspace*{0.0cm}
\end{center}
\begin{footnotesize}
Note:  Posterior mean of model fitted  rainfall.  
\end{footnotesize}
\end{figure}

In order to show the intense variability in the rainfall pattern in the Northeast region, we present in Figure \ref{prevpluv} the model adjustment for the third quarter of 2001, when there was severe drought. It is possible to observe that in this quarter a significant part of the Northeast region had very low precipitation, especially for the semi-arid portion, with precipitation near zero in this period.

\section{Extensions\label{extensions}}

In this section, we present some possible extensions of the use and specification of the method proposed here. The first is the construction of the out-of-sample forecasts, and the second is the use of non-stationary spatial covariance matrices. 

\subsection{Forecasting}

The preparation of climate forecasts is possibly the best-known area of meteorology, since it has practical impacts for society as whole. At the same time, it is possibly the most complex prediction problem, since accurate climate forecasting involves nonlinear computational models of immense complexity and a high number of determining factors. In this way, the idea of obtaining accurate forecasts using simple models is quite naive. However, since we are working with quarterly data, a significant portion of the sources of variation is eliminated, so we can see how the model behaves to forecast the overall behavior. 

\begin{figure}
\caption{ \textbf{Forecast - Average Temperature - 2014/1}\label{forecastpluv}}
\begin{center}
\includegraphics[width=90mm,angle=270]{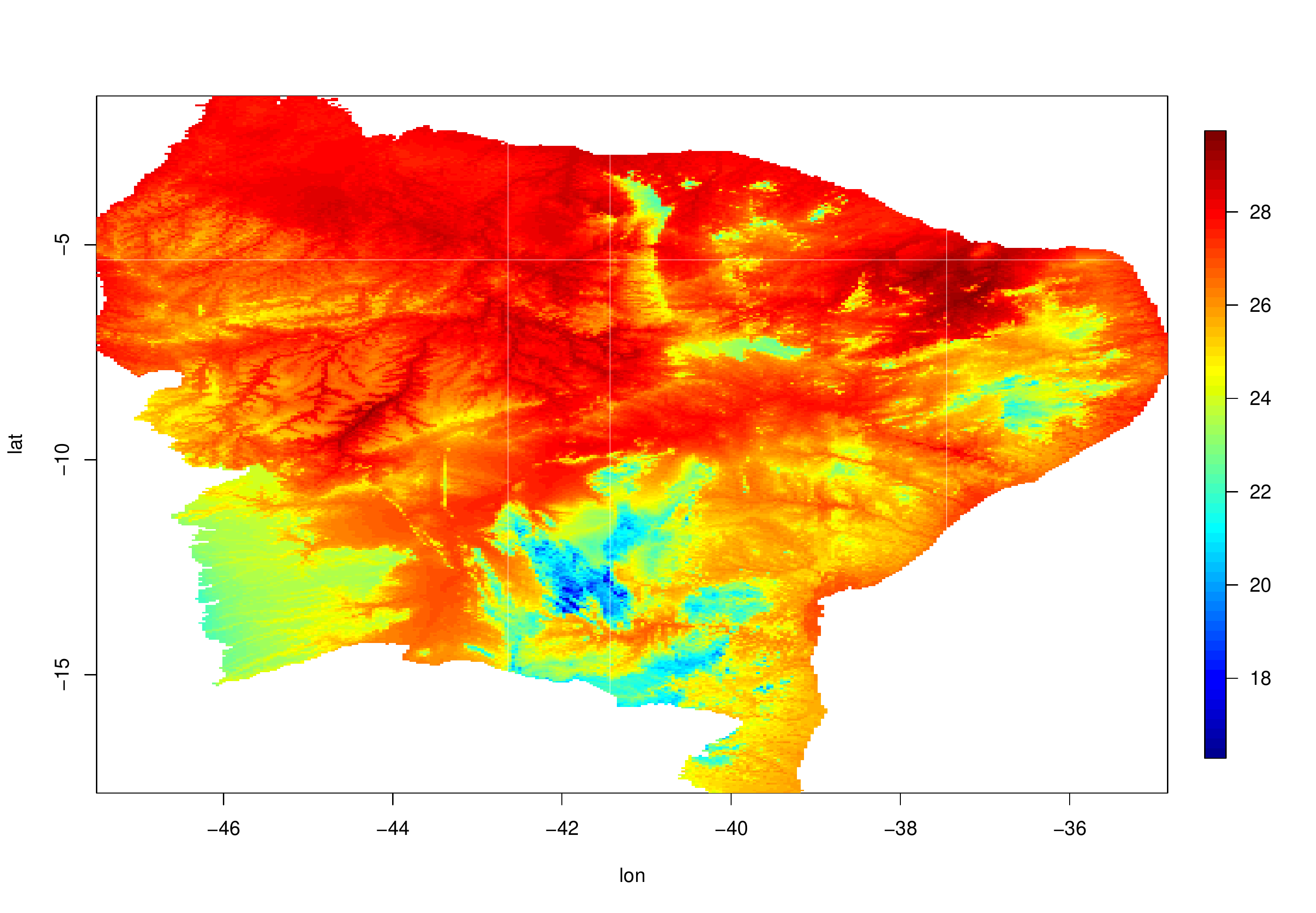}
\vspace*{0.0cm}
\end{center}
\begin{footnotesize}
Note:  Posterior mean of model forecasts.  
\end{footnotesize}
\end{figure}

We performed a pseudo-out-of-sample preview exercise to verify this possibility. We estimated the model for average temperature using data available from 1961 to 2013, and we forecasted from 1 to 4 quarters ahead using the model. Analogous to the model fit construction, we performed the prediction using the 1 to 4 step-ahead forecast for the trend, seasonality and cycle components, and adding  the estimated spatial random effect component. Figure \ref{forecastpluv} shows the forecast for the first quarter of 2014.

We performed a simple predictive performance procedure, comparing predicted values for the model with the values observed by the monitoring stations in the 2014/1 - 2014/4 quarters. Table 4 presents the results of this analysis, separated for each quarter. Although we are not comparing the prediction results with other models, it is possible to observe that the model presents relatively low values in measures such as average prediction error (ME), mean absolute error (MAE), mean percent error (MPE) and mean absolute percentage error (MAPE). An informal comparison can be made with the results shown in Table 1 of  \cite{Beenstock20161234}, which shows the performance measures of 22 climate change forecasting models. The results show good performance of the model used in this work, although again the comparison is informal since \cite{Beenstock20161234} analyze other datasets and data frequencies.

\begin{table}[ht]
\caption{\textbf{Forecast Measures by Quarter}} 
\centering
\begin{tabular}{rrrrrrrr}
  \hline
 & ME & RMSE & MAE & MPE & MAPE & ACF1 & Theil's U \\ 
  \hline
  2014/1 & 0.055 & 0.942 & 0.723 & 0.204 & 2.732 & -0.075 & 0.362 \\ 
  2014/2 & -0.184 & 1.029 & 0.860 & -0.794 & 3.357 & -0.088 & 0.387 \\ 
  2014/3 & -0.375 & 1.475 & 1.169 & -1.299 & 4.833 & -0.155 & 0.548 \\ 
  2014/4 & 0.176 & 1.500 & 1.216 & 0.827 & 4.611 & 0.000 & 0.574 \\ 
   \hline
\end{tabular}
\label{forecaststat}
\end{table}

\subsection{Non-stationary covariance structures\label{nonstatsec}}

A possible limitation of the method used so far to measure climate change patterns is the fact that the spatial covariance structure is assumed to be spatially stationary, depending on the parameterization of Matérn covariance with two fixed parameters throughout the sample. Although this structure may be adequate in many contexts and replicate the pattern of observed spatial heterogeneity, one possibility is the use of non-stationary spatial covariance structures to characterize processes of spatial dependence, as discussed for example in \cite{Ingebrigtsen201420} and the other references cited in this paper. The spatial non-stationarity would be characterized by a spatial dependence pattern that is a function of the location, not just the distances between locations, analogous to the problem of non-stationarity in time series. In this case, the non-stationarity could capture distinct dependency patterns for regions with distinct conditioning characteristics.

One possible way to introduce non-stationarity into spatial models is to make the covariance function a function of some explanatory variable, as used in \cite{Schmidt2011} and \cite{Ingebrigtsen201420}. In particular, \cite{Ingebrigtsen201420} use altitude as an explanatory variable in the spatial covariance function, employing the method proposed in \cite{spde} to perform regression modeling of annual rainfall indices in Norway. The proposed modification in \cite{Ingebrigtsen201420} to introduce non-stationarity into the covariance function is to make the parameters that characterize the spatial covariance function,   log $\tau$ and log $\kappa$,  functions of some feature associated with each location in space. The formulation proposed by \cite{Ingebrigtsen201420} uses a basis expansion for log $\tau$ and log $\kappa$ in the form:

\begin{equation}
\begin{array}{c}
\log \tau(h)=\theta_1^\tau+\sum_{k=2}^N b_k^\tau (h) \theta_k^\tau\\
\log \kappa(h)=\theta_1^\kappa+\sum_{k=2}^N b_k^\kappa (h) \theta_k^\kappa\\
\end{array}\label{modelnon}
\end{equation}

\noindent where the $\theta$ coefficients are weighting parameters and $b_k^\tau (h)$  and  $b_k^\kappa (h)$ are  deterministic basis expansions at the nodes of the mesh triangulation used. This procedure is equivalent to a modification in the precision matrix used to represent the Gaussian Markov random field associated with the spde solution; \cite{Ingebrigtsen201420} discuss the details and modifications needed in spatial representation to contemplate this structure. Note that the stationary case reduces to  the constraints log $\tau(h)=\theta_1^\tau$   and  log $\kappa=\theta_1^\kappa$. 

We introduce a non-stationary structure in our work by making the parameter log $\tau$  a linear function of the altitude in the estimation of the model for average temperature, which is equivalent to introducing an additional parameter $ \theta_k^\tau$  in the model, capturing the effect of altitude on the covariance function. We  tested other specifications with both log $\tau $ and log $\kappa$ variants, but the specification with only log $\tau$ variant performed best in terms of DIC.

\begin{table}
\protect\caption{\label{tab:meannon} \textbf{Estimated parameters - Average temperature with non-stationary spatial covariance}}
\begin{small}
\begin{tabular}{ccccccc}
\hline 
 & {\small{}mean} & {\small{}sd} & {\small{}.025q} & {\small{}.5q} & {\small{}.975q} & {\small{}mode}\tabularnewline
\hline 
{\small{}Altitude} &   -0.0075& 0.0002&    -0.0080&  -0.0075 &   -0.0070& -0.0075  \tabularnewline
{\small{}Latitude } & 0.1309& 0.0338   &  0.0645 &  0.1309  &   0.1972 & 0.1309   \tabularnewline
{\small{}Distance to Sea } &  0.0048 & 0.0006  &   0.0035 &  0.0048  &   0.0061 & 0.0048 \tabularnewline
Precision Gaussian &    0.9655& 1.250e-02    & 0.9413 &    0.9655 &    0.9902  &   0.9653\tabularnewline
Precision RW &923.6432 &6.258e+02 &  212.4327 &  769.3595 & 2543.7127 &  515.0026\tabularnewline
Precision Seasonal  &   29366.9054& 2.368e+04 & 3700.2466 &23241.2592 &90827.7800 &10744.9796\tabularnewline
Precision Cycle  &5.4977& 7.176e-01  &   4.1725 &    5.4733&     6.9862 &    5.4445\tabularnewline
PACF1  &  0.4006 &7.210e-02  &   0.2617 &    0.3992 &    0.5433 &    0.3923\tabularnewline
PACF2 &   -0.0625& 8.150e-02  &  -0.2110 &   -0.0670  &   0.1074  &  -0.0803\tabularnewline
log $\tau$ &     -1.1129& 1.794e-01 &   -1.4616 &   -1.1142&    -0.7583 &   -1.1178\tabularnewline
log $\kappa$ &     0.2743& 1.903e-01  &  -0.1135   &  0.2801  &   0.6340  &   0.2968\tabularnewline
 $\theta_k^\tau$ &    -0.00527& 3.480e-04 &   -0.01185  &  -0.00538  &   0.00179  &  -0.00569\tabularnewline
Marginal Lik. &-18167.66   &  obs &  12323 &  &  & \tabularnewline
DIC  &35674.93    &   &  &  &  & \tabularnewline
\hline 
\end{tabular}
\end{small}
\begin{tiny}
\end{tiny}
\end{table}

\begin{figure}
\caption{ \textbf{Spatial random effects  - Average temperature with non-stationary spatial covariance}\label{retempnon}}
\begin{center}
\includegraphics[width=90mm,angle=270]{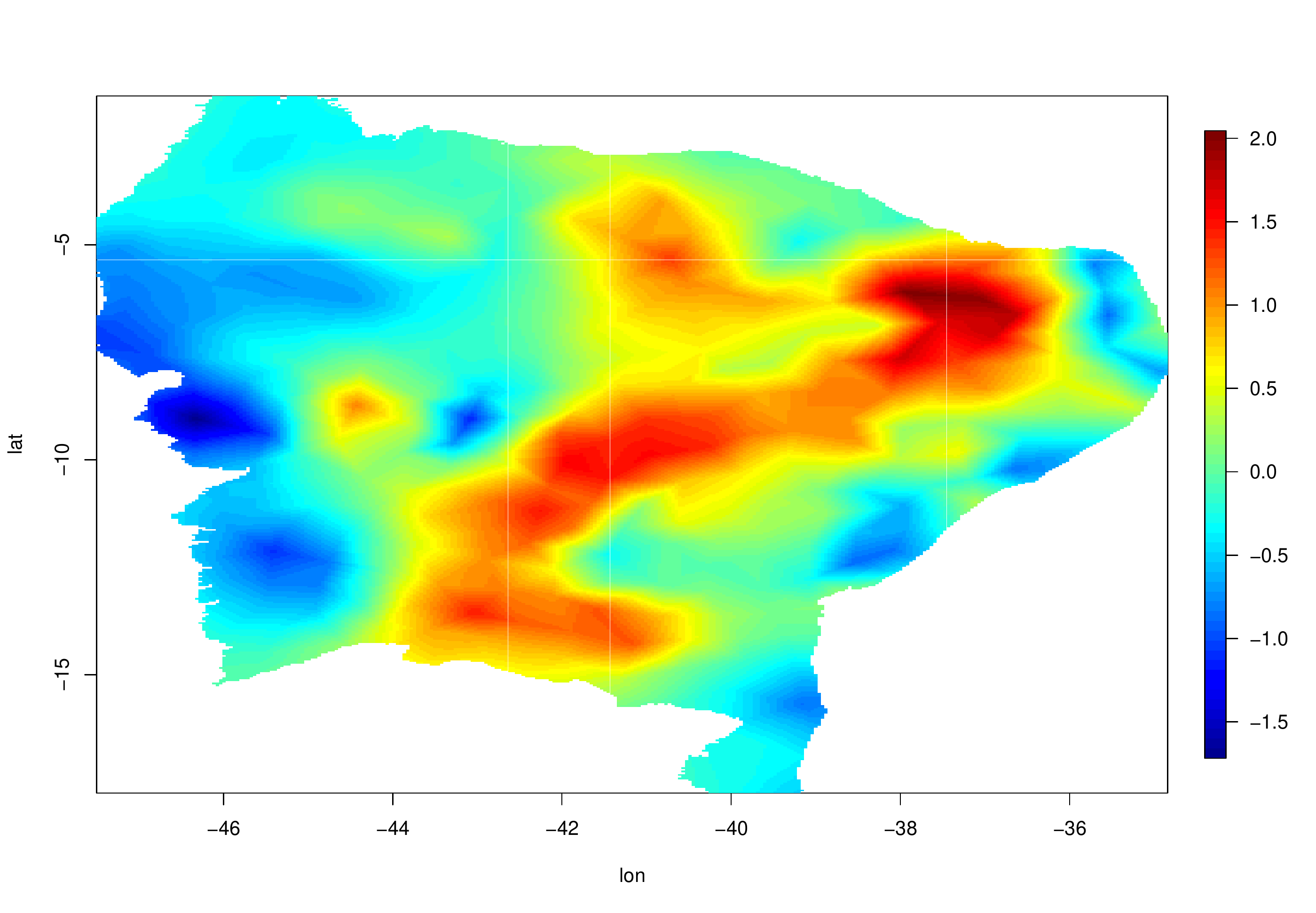}\\
\vspace*{0.0cm}
\end{center}
\begin{footnotesize}
Note:  Posterior mean of estimated spatial random effects. 
\end{footnotesize}
\end{figure}

Table \ref{tab:meannon} presents the results of the estimation with the spatially non-stationary model. The DIC of the model with the non-stationary spatial covariance function is 35674.93 against the value of  35676.84 for the standard stationary model, representing a marginal improvement. In general, the results in terms of trend, seasonality and cycle extraction are very similar to those obtained with the stationary model, and therefore are not shown. Figure \ref{retempnon} shows the spatial random effects estimated by the non-stationary formulation. These are very similar to those obtained by the stationary model. Although in the present problem the use of this particular structure to introduce a non-stationary formulation for the spatial covariance function did not significantly alter the results obtained, it is important to note that this formulation may be important in other climate related problems, and therefore is mentioned in this paper.

\section{Conclusions\label{conclusions}}

In addition to the possible problems caused to the environment by climate change processes, the economic impacts of these changes are also very important. \cite{Nordhaus2016} discusses the economic cost of the increase observed in global temperatures, and shows that the economic costs associated with the current patterns of climate change will be relevant even with the immediate adoption of effective policies to combat climate change. Using a long run risk model, \cite{Bansal2016}  show there is a high welfare cost associated with these climate changes. Thus, the measurement of climate change patterns is a social and economic problem of the first order.

The present work contributes to this problem with a new method for the time series decomposition into trend, seasonality and cycle components that allows incorporating the important characteristics of climatic data, such as the variability observed in a large number of measuring stations, the missing data problem and the spatial heterogeneity of these series. The incorporation of these effects into the estimation of unobserved components is essential for accurate measurement of the effects of climate change. If spatial effects are not incorporated into the model, the components may not be correctly estimated, leading to imprecise inferences about the true components of trend, seasonality, and cycle in these series, as shown in Section 4. Similarly, we show the importance of including the multiple sources of data from the various weather monitoring stations for a correct estimation of the uncertainty associated with the patterns of climate change. 

In addition to capturing the patterns of climate change, this method is a useful contribution to climate analysis by allowing the estimation of spatial heterogeneity through a spatial random effects structure projected in the spatial continuum, controlling for the dynamic effects existing in these series, generalizing the method of \cite{spde}, which does not include dynamic components. We also show some possible ways to use and generalize this method, such as the construction of forecasts and the possible inclusion of non-stationary components in the spatial covariance function. 

The results give support to the increase in the trend component of the observed temperatures for the Northeast region of Brazil, compatible with the evidence of permanent increases in the temperatures observed globally  (e.g., \cite{Ji2014}), and supporting the hypothesis of climate change. This application shows that the method allows including and capturing the richness and climatic amplitude observed in the modeling of climatic patterns, indicating that econometric methods can be used to perform complex analyses in climatology, complementing the computational simulation models typically employed in these analyses. 

\bibliographystyle{chicago}
\bibliography{fineconometricsfullc}

\end{document}